\documentclass[british,english]{article}
\usepackage[T1]{fontenc}
\usepackage[latin9]{inputenc}
\usepackage{geometry}
\usepackage{babel}
\usepackage{amsmath}
\usepackage{amssymb}
\usepackage{graphicx}
\usepackage[authoryear]{natbib}
\usepackage[unicode=true,pdfusetitle,
 bookmarks=true,bookmarksnumbered=false,bookmarksopen=false,
 breaklinks=false,pdfborder={0 0 1},backref=false,colorlinks=false]
 {hyperref}

\makeatletter

\providecommand{\tabularnewline}{\\}

\usepackage{hyperref}
\usepackage[linesnumbered,ruled]{algorithm2e}
\usepackage{authblk}

\@ifundefined{showcaptionsetup}{}{%
 \PassOptionsToPackage{caption=false}{subfig}}
\usepackage{subfig}
\makeatother

\begin{document}
\title{Ensemble Kalman inversion approximate Bayesian computation}
\author[1,2]{Richard G. Everitt}
\affil[1]{Department of Statistics, University of Warwick}
\affil[2]{The Zeeman Institute for Systems Biology \& Infectious Disease Epidemiology Research, University of Warwick}
\affil{Email: richard.everitt@warwick.ac.uk}
\maketitle
\begin{abstract}
Approximate Bayesian computation (ABC) is the most popular approach
to inferring parameters in the case where the data model is specified
in the form of a simulator. It is not possible to directly implement
standard Monte Carlo methods for inference in such a model, due to
the likelihood not being available to evaluate pointwise. The main
idea of ABC is to perform inference on an alternative model with an
approximate likelihood (the \emph{ABC likelihood}), estimated at each
iteration from points simulated from the data model. The central challenge
of ABC is then to trade-off bias (introduced by approximating the
model) with the variance introduced by estimating the ABC likelihood.
Stabilising the variance of the ABC likelihood requires a computational
cost that is exponential in the dimension of the data, thus the most
common approach to reducing variance is to perform inference conditional
on summary statistics. In this paper we introduce a new approach to
estimating the ABC likelihood: using iterative ensemble Kalman inversion
(IEnKI) \citep{iglesias_regularizing_2016,iglesias_bayesian_2018}.
We first introduce new estimators of the marginal likelihood in the
case of a Gaussian data model using the IEnKI output, then show how
this may be used in ABC. Performance is illustrated on the Lotka-Volterra
model, where we observe substantial improvements over standard ABC
and other commonly-used approaches.
\end{abstract}

\section{Introduction}

\subsection{Intractable likelihoods and simulation-based inference}

This paper considers the problem of Bayesian inference for parameters
$\theta$ of a data generating process $P\left(\cdot\mid\theta\right)$
used as a model for data $y_{\text{obs}}$, using a prior $p$ on
$\theta$. We use $\pi$ to denote the posterior distribution of $\theta$
given $y_{\text{obs}}$, with $\tilde{\pi}$ denoting the unnormalised
posterior given by $\tilde{\pi}\left(\theta\mid y_{\text{obs}}\right)=p(\theta)P(y_{\text{obs}}\mid\theta)$.
We consider the case where the likelihood $P(y_{\text{obs}}\mid\theta)$
is intractable in that it cannot be evaluated pointwise at $\theta$.
There are several situations in which this can occur:
\begin{enumerate}
\item in the case of ``tall'' data \citep{Bardenet2017}, where the $P$
consists of a product of a very large number of terms;
\item where $\pi$ is ``doubly intractable'' \citep{Murray2006}, referring
to $P$ having an intractable normalising constant;
\item where latent variables $x$ are involved in the specification of $P$
as follows
\begin{equation}
P(y\mid\theta)=\int_{x}g(y\mid x,\theta)f(x\mid\theta)dx;\label{eq:latent}
\end{equation}
\item where $P$ is specified by a piece of software that can be simulated
- it is possible to simulate $y\sim P\left(\cdot\mid\theta\right)$
- but it is impractical to write out $P$ in mathematical form.
\end{enumerate}
These cases overlap: for example usually in situation 4, it would
in principle be possible to write the model as specified in equation
(\ref{eq:latent}), were the details of the software implementation
available. Inference in situation 4 has become known as \emph{simulation-based
inference} \citep{cranmer_frontier_2020-1}, a term given to collectively
describe methods that approximate the posterior when $P$ can be simulated,
but not evaluated pointwise at $\theta$.

The fundamental SBI technique is approximate Bayesian computation
(ABC; \citet{tavare_inferring_1997}). ABC approximates the likelihood
using simulations from $P$, based on the simple principle that ``good''
parameters are more likely to generate simulations $y$ that are close
to the observed data $y_{\text{obs}}$. For each $\theta$, $M$ simulations
$\left\{ y^{j}\right\} _{j=1}^{M}$ are taken from $P\left(\cdot\mid\theta\right)$,
summary statistics $s^{j}=S(y^{j})$ of each $y^{j}$ are calculated,
then a kernel $K_{\varepsilon}\left(s_{\text{obs}}\mid s^{j}\right)$
(where $s_{\text{obs}}=S(y_{\text{obs}})$) is calculated for each
$s^{j}$. The likelihood is then approximated using
\begin{equation}
\frac{1}{M}\sum_{j=1}^{M}K_{\varepsilon}\left(s_{\text{obs}}\mid s^{j}\right),\label{eq:est_abc_llhd}
\end{equation}
this being a Monte Carlo approximation of the \emph{ABC likelihood}
\begin{equation}
\int_{s}K_{\varepsilon}\left(s_{\text{obs}}\mid s\right)P_{s}\left(s\mid\theta\right)ds,\label{eq:abc_llhd}
\end{equation}
where $P_{s}$ is the distribution resulting from the transformation
$s=S(y)$ of $y\sim P(\cdot\mid\theta)$. The kernel $K_{\varepsilon}$
takes values in $[0,\infty)$ and is chosen to take higher values
the closer $s^{j}$ is to $s_{\text{obs}}$, with the ``tolerance''
$\varepsilon>0$ controlling the ``width'' of the kernel. Popular choices
for $K_{\varepsilon}$ in include:
\begin{itemize}
\item Uniform
\[
K_{\varepsilon}\left(s_{\text{obs}}\mid s\right)=\begin{cases}
1 & \text{if }d(s_{\text{obs}},s)<\varepsilon\\
0 & \text{otherwise}
\end{cases},
\]
where $d$ is some distance metric. A common choice is the weighted
Euclidean distance
\[
d(s_{\text{obs}},s)=\left(\sum_{i=1}^{d_{s}}\left(\frac{s_{\text{obs},i}-s_{i}}{\sigma_{i}}\right)^{2}\right)^{1/2},
\]
where $d_{s}$ is the dimension of $s$ and each $\sigma_{i}$ is
chosen such that each of the terms in the sum have a similar scale,
so that all summary statistics have a comparable influence.
\item Gaussian
\[
K_{\varepsilon}\left(s_{\text{obs}}\mid s\right)=\left(2\pi\varepsilon\right)^{-d_{y}/2}\prod_{i=1}^{d_{s}}\frac{1}{\sigma_{i}^{2}}\exp\left(-\frac{1}{2\varepsilon^{2}}\sum_{i=1}^{d_{s}}\left(\frac{s_{\text{obs},i}-s_{i}}{\sigma_{i}}\right)^{2}\right),
\]
i.e. the multivariate Gaussian distribution with mean $s$ and covariance
$\varepsilon^{2}\Sigma_{s}$ where $\Sigma_{s}$ is diagonal with $i$th
diagonal element $\sigma_{i}^{2}$, similarly to scale the summary
statistics.
\end{itemize}
The reduction of the data $y$ to the summary $s$ means that the
ABC posterior (the posterior using the ABC likelihood) is a posterior
conditional on $s_{\text{obs }}$ rather than $y_{\text{obs}}$ which
represents a loss of informaton about $\theta$ unless $s$ is sufficient.
As $\varepsilon\rightarrow0$, the ABC posterior converges to the true
posterior conditional on $s_{\text{obs }}$. Using $\varepsilon>0$,
which is almost always required in practice, introduces a further
approximation. Most other approaches to SBI also rely on approximations
of some kind (see, e.g. \citet{cranmer_frontier_2020-1}).

The estimated ABC likelihood in equation (\ref{eq:est_abc_llhd})
is usually used within a Monte Carlo method that explores the space
of $\theta$, for example within a Markov chain Monte Carlo (MCMC)
algorithm. The resultant algorithm \citep{marjoram_markov_2003-2}
is known as ABC-MCMC, which can be seen to be an instance of a pseudo-marginal
\citep{Beaumont2003,Andrieu2009} algorithm. It is well-established
(e.g. \citet{Andrieu2014a}) that for a pseudo-marginal algorithm
to be efficienct, the estimated marginal likelihood needs to have
a relatively small variance.

The past 25 years have seen the development of a range of alternative
ABC and other SBI methods. A core challenge is that of performing
SBI in ``high'' dimensions: most approaches rely on $d_{s}$ being
small. In particular, the variance of the likelihood estimator in
equation (\ref{eq:est_abc_llhd}) requires $M$ to be exponential
in $d_{s}$ to stablise its variance. This also limits the use of
ABC to low-dimensional parameter spaces, since a dimension $d_{\theta}$
of $\theta$ larger than $d_{s}$ results in identifiability problems
in the posterior.

\subsection{Alternatives and extensions to ABC}

For the intractable likelihood in situation 3, SBI is often not the
most effective choice of method. Whilst the integral $P$ may be intractable,
it is usually possible to evaluate the functions $f$ and $g$ in
equation (\ref{eq:latent}), so we may use Monte Carlo approaches
whose target is the joint posterior $\pi(\theta,x\mid y_{\text{obs}})$,
then use the $\theta$-points to approximate $\pi(\theta\mid y_{\text{obs}})$.
One such approach is particle MCMC \citep{Andrieu2010c}, which uses
sequential Monte Carlo (SMC) methods to explore $x$-space conditional
on $\theta$. In situation 3, such an approach might be seen to be
favourable to SBI, since it has the does not suffer the approximations
introduced in SBI. Indeed, some recent work \citep{Prangle2016,kerama_rare_2022,zhang_improvements_2022}
has focussed on using a reparameterisation to cast models that might
usually be tackled using ABC in the form in equation (\ref{eq:latent}).
Specifically, the idea is to rewrite the simulator model $P$ as a
deterministic transformation $G\left(\theta,u\right)$ of the parameter
$\theta$ and random variables $u$ whose prior $\phi\left(u\mid\theta\right)$
is a normalised tractable distribution. The ABC likelihood then changes
from equation (\ref{eq:abc_llhd}) to
\begin{equation}
\int_{u}K_{\varepsilon}\left(s_{\text{obs}}\mid S(G(\theta,u))\right)\phi(u\mid\theta)du.\label{eq:re_abc_llhd}
\end{equation}
We refer to this situation as \emph{rare event }(RE) ABC after the
interpretation of the estimation of equation (\ref{eq:re_abc_llhd})
as rare event estimation. Although there are situations where using
such a reparameterisation may be difficult (for example when given
a simulator in the form of legacy code), the significant advantage
over standard ABC is that, now that the intractable $P$ does not
appear in the expression, we are free to use alternative Monte Carlo
methods to sample from the target distribution. For example, particle
MCMC is used in \citet{Prangle2016}, referred to as RE-ABC-MCMC,
using SMC to explore $u$-space conditional on $\theta$. The SMC
yields a lower variance estimator than equation (\ref{eq:est_abc_llhd}),
hence usually yields a substantially more efficient MCMC chain compared
to standard ABC-MCMC.

\subsection{Proposed approach}

Despite the advances described in the preceding section, there remain
many situations where ABC might be a more efficient approach than,
say, particle MCMC, when in situation 3. When the dimension $d_{x}$
of $x$-space is high, or $x$-space has a complex structure, using
Monte Carlo methods to explore this space may be inefficient: we may
need a very large computational time to obtain low variance Monte
Carlo estimates. Examples include: the situation for which ABC was
originally designed in population genetics, where $x$ is a tree describing
the ancestry of a number of individuals \citep{tavare_inferring_1997};
or time-series data with a large number of noisy observations, where
$x$ is a latent time series (e.g. \citet{Andrieu2010c}). The same
issue is present in rare event ABC when $d_{u}$, the dimension of
$u$, is high. ABC largely avoids this issue by using a likelihood
that involves integration over space that is typically lower dimension:
over dimension $d_{s}$ (equation (\ref{eq:abc_llhd})) compared to
$d_{x}$ (equation (\ref{eq:latent})). Therefore, despite the bias
present in an ABC posterior, the lower variance may result in smaller
errors given a fixed computational effort.

In this paper we propose a new approach to estimating the ABC likelihood
in equation (\ref{eq:abc_llhd}) using iterative ensemble Kalman inversion
(IEnKI) \citep{iglesias_regularizing_2016,iglesias_bayesian_2018},
and use this with an MCMC algorithm. IEnKI is a Monte Carlo approach
that has some similarities with SMC, so in this sense our new approach
has the flavour of RE-ABC-MCMC. However, it has the crucial difference
that rather than using the form of the ABC likelihood in equation
(\ref{eq:re_abc_llhd}) which integrates over a space of dimension
$d_{u}$, it uses the original form in equation (\ref{eq:abc_llhd})
which integrates over a space of dimension $d_{s}$. We show empirically
that this estimator can have substantially lower variance than the
standard estimated ABC likelihood in equation (\ref{eq:est_abc_llhd}),
and that the variance can be relatively low when $d_{s}$ is much
larger than is possible for standard ABC. We show the new approach
compares favourably with a number of alternative approaches on the
Lotka-Volterra model, including standard ABC, synthetic likelihood
and particle MCMC.

The structure of the paper is as follows. In section \ref{sec:Iterative-ensemble-Kalman}
we recap IEnKI and introduce new marginal likelihood estimators using
the output of this algorithm. In section \ref{sec:IEnKI-ABC} we show
how to use the new marginal likelihood estimators to estimate the
ABC likelihood, highlighting some properties of the new IEnKI that
facilitate its use for ABC. We illustrate the performance of the new
methods in section \ref{sec:Applications}, before a discussion in
section \ref{sec:Discussion}.

\section{Iterative ensemble Kalman inversion\label{sec:Iterative-ensemble-Kalman}}

In this section we describe IEnKI, then outline how a marginal likelihood
estimator may be devised from the output of this algorithm.

\subsection{Algorithm}

\subsubsection{SMC samplers}

The structure of IEnKI follows that of SMC samplers \citep{DelMoral2006c}:
a method designed to produce a Monte Carlo approximation to a sequence
of targets. We begin by introducing notation. Let $\pi_{t}$ (for
$t=0:T$) be a sequence of target distributions, with $\tilde{\pi}_{t}$
denoting an unnormalised form of each distribution and $Z_{t}$ its
normalising constant. SMC samplers are commonly used for approximating
Bayesian posterior distributions, so a popular sequence of targets
is
\begin{equation}
\tilde{\pi}_{t}(x)=p(x)l^{\alpha_{t}}(x),\label{eq:annealed_target}
\end{equation}
where $p$ is a prior distribution and $l$ a likelihood, and $0=\alpha_{0}<...<\alpha_{T}=1$.
This sequence has several appealing properties: the first distribution
is equal to the prior and is usually easy to simulate; the final distribution
is equal to the posterior; and the sequence of $\alpha_{t}$ can be
chosen so as to create a path of distributions between the prior and
posterior such that the approximation to the target $\pi_{t-1}$ may
be used (via the SMC algorithm) to constuct an approximation to $\pi_{t}$.

The output of an SMC sampler is a population of weighted particles
$\{(x_{t}^{j},w_{t}^{j})\}_{j=1}^{M}$ at each iteration that is used
to approximate $\pi_{t}$ and estimate the normalising constant $Z_{t}$.
Here we present the most commonly-used case, where at iteration $t$
each particle is moved using $K_{t}$, an MCMC kernel with invariant
distribution $\pi_{t}$ At iteration $t$ the unnormalised weight
of each particle is calculated using $\tilde{w}_{t}^{j}=l^{\alpha_{t}-\alpha_{t-1}}(x_{t-1}^{j})$,
then the unnormalised weights $\{\tilde{w}_{t}^{j}\}_{j=1}^{M}$ are
normalised to give $\{w_{t}^{j}\}_{j=1}^{M}$ and the new particle
population is simulated from $\sum_{j=1}^{M}w_{t}^{j}K_{t}(\cdot\mid x_{t-1}^{j})$.
The performance of estimators from SMC samplers is critically dependent
on choices made in the design of the algorithm, namely:
\begin{enumerate}
\item The choice of the the sequence $(\alpha_{t})$. If $\alpha_{t}$ is
too far from $\alpha_{t-1}$ the algorithm will become \emph{degenerate},
where one particle has most of the weight, resulting in high-variance
estimators. If $\alpha_{t}$ is too close to $\alpha_{t-1}$, the
computational cost of the algorithm will be high. It is common to
automatically construct the sequence $(\alpha_{t})$ adaptively as
the algorithm is running, where $\alpha_{t}$ is chosen at each iteration
such that degeneracy is avoided \citep{DelMoral2012g}.
\item The choice of MCMC kernel $K_{t}$. The kernel needs to move efficiently
around the space, with the critical point that $K_{t}$ does indeed
need to change with $t$, since a kernel that is efficient in exploring
target $\pi_{1}$ is usually not efficient at exploring $\pi_{T}$.
Adaptive schemes are commonly used, for example choosing the scale
of a random walk proposal in a Metropolis-Hastings move at iteration
$t$ based on a scale estimated from the current particle population.
\end{enumerate}

\subsubsection{Iterative ensemble Kalman inversion\label{subsec:Iterative-ensemble-Kalman}}

Whilst adaptive schemes can reduce the burden on the user of designing
the SMC sampler, in some applications the implemention of an effective
algorithm can still be non-trivial and the computational cost of the
method prohibitively high. IEnKI was designed with the motivation
of avoiding these issues. 

The starting point in our description of the algorithm is in using
the same sequence of targets as in equation (\ref{eq:annealed_target}),
but with the assumption that the prior $p$ is Gaussian and that the
data model $g$ is also Gaussian, i.e.
\[
p(x)=\left(2\pi\right)^{-d_{x}/2}\det\left(\Sigma_{0}\right)^{-1/2}\exp\left(-\frac{1}{2}\left(x-\mu_{0}\right)^{T}\Sigma_{0}^{-1}\left(x-\mu_{0}\right)\right)
\]
\[
g\left(y_{\text{obs}}\mid x,\Sigma_{y}\right)=\left(2\pi\right)^{-d_{y}/2}\det\left(\Sigma_{y}\right)^{-1/2}\exp\left(-\frac{1}{2}\left(y_{\mbox{obs}}-H(x)\right)^{T}\Sigma_{y}^{-1}\left(y_{\mbox{obs}}-H(x)\right)\right),
\]
where $H:\mathbb{R}^{d_{x}}\rightarrow\mathbb{R}^{d_{y}}$. Let $l(x):=g\left(y_{\text{obs}}\mid x,\Sigma_{y}\right)$
be the likelihood of $x$ and let $\tilde{\pi}_{t}(x)=p(x)l^{\alpha_{t}}(x)$
be the sequence of unnormalised targets used as in SMC, each with
normalising constant $Z_{t}$. This choice results in the recursive
relationship $\tilde{\pi}_{t}(x)=\tilde{\pi}_{t-1}(x)l^{\alpha_{t}-\alpha_{t-1}}(x)$.
Simplifying notation by taking $\gamma_{t}:=\left(\alpha_{t}-\alpha_{t-1}\right)^{-1}$
we have
\[
l^{\alpha_{t}-\alpha_{t-1}}\left(x\right)=\left(2\pi\right)^{-\frac{d_{y}}{2\gamma_{t}}}\det\left(\Sigma_{y}\right)^{-\frac{1}{2\gamma_{t}}}\exp\left(-\frac{1}{2\gamma_{t}}\left(y_{\mbox{obs}}-H(x)\right)^{T}\Sigma_{y}^{-1}\left(y_{\mbox{obs}}-H(x)\right)\right).
\]
Note that $l^{\gamma_{t}}\left(x\right)$ is proportional to the Gaussian
distribution with covariance $\gamma_{t}\Sigma_{y}$. Temporarily,
for the purposes of exposition, we assume that $H$ is a linear function
and switch the notation so that $H\in\mathbb{R}^{d_{x}\times d_{y}}$.
Since $p$ is Gaussian, this observation makes it clear that every
member of sequence of normalised target distributions $\pi_{t}$ is
also Gaussian, and a conjugate update may be used to find $\pi_{t}(x)$
from $\pi_{t-1}(x)$. Let $\mu_{t}^{x}$ and $C_{t}^{xx}$ respectively
be the mean and covariance of $\pi_{t}$. We then have, following
the Kalman filter update equations:
\[
S_{t}=HC_{t-1}^{xx}H^{T}+\gamma_{t}\Sigma_{y}
\]
\[
K_{t}=C_{t-1}^{xx}H^{T}S_{t}^{-1}
\]

\[
\mu_{t}^{x}=\mu_{t-1}^{x}+K_{t}\left(y_{\text{obs}}-H\mu_{t-1}^{x}\right)
\]

\[
C_{t}^{xx}=\left(I_{d_{x}}-K_{t}H\right)C_{t-1}^{xx}.
\]

When the Gaussian assumptions hold, there is clearly no need to use
such a recursive algorithm to calculate $\pi_{t}$: the same result
would be achieved through a single conjugate update from $p(x)$ to
$p(x)l(x)$. The interest in this formulation arises when the sequence
of targets is non-Gaussian. In place of the exact update, ensemble
Kalman methods use a Monte Carlo approximation of the update: a population
of ensemble members $\{x_{t}^{j}\}_{j=1}^{M}$ is used to represent
the target $\pi_{t}$, and at each iteration these points are updated
such that their sample mean and variance estimate $\mu_{t}^{x}$ and
$C_{t}^{xx}$. Whilst convergence to the exact updates is guaranteed
for the Gaussian case \citep{mandel_convergence_2011}, ensemble Kalman
approaches are frequently used outside of this setting, as an approximate
approach where convergence is not guaranteed \citep{katzfuss_understanding_2016,roth_ensemble_2017}.
IEnKI is the name given to the method that applies the ensemble Kalman
idea to the sequence of targets described in this section. 

We describe this approach precisely in algorithm \ref{alg:senki},
where we revert back to the non-linear/Gaussian case, taking $H:\mathbb{R}^{d_{x}}\rightarrow\mathbb{R}^{d_{y}}$.
To simplify notation we use the notation $h_{t}^{(j)}:=H(x_{t}^{(j)})$,
using $\mu^{h}$ and $C^{hh}$ to denote respectively the mean and
covariance of $h$, and $C^{xh}$ to denote the cross-covariance of
$x$ and $h$.

\begin{algorithm} \caption{Stochastic iterative ensemble Kalman inversion.}\label{alg:senki}

Simulate $M$ points, $\left\{ x_{0}^{j} \right\}_{j=1}^{M} \sim p$;\

For $j=1:M$, let $h^{j}_{0} = H(x^{j}_{0})$;

$\hat{\mu}^{x}_{0} = \frac{1}{M} \sum_{j=1}^M x^{j}_{0}$;
	
$\hat{\mu}^{h}_{0} = \frac{1}{M} \sum_{j=1}^M h^{j}_{0}$;

$\hat{C}^{x h}_{0} = \frac{1}{M-1} \sum_{j=1}^M \left( x^{j}_{0} - \hat{\mu}^{x}_{0} \right) \left( h^{j}_{0} - \hat{\mu}^{h}_{0} \right)^T$;

$\hat{C}^{h h}_{0} = \frac{1}{M-1} \sum_{j=1}^M \left( h^{j}_{0} - \hat{\mu}^{h}_{0} \right) \left( h^{j}_{0} - \hat{\mu}^{h}_{0} \right)^T$;

\For {$t=1:T$}
{

	$\hat{K}_{t} = \hat{C}^{x h}_{t-1} \left( \hat{C}^{h h}_{t-1} + \gamma_t \Sigma_y \right)^{-1}$;

	\For{$n=1:N_a$}
	{
		Simulate $\tilde{y}^{j}_{t} \sim \mathcal{N}\left( h^{j}_{t-1}, \gamma_t \Sigma_y \right)$;

		Shift $x^{j}_{t} = x^{j}_{t-1} + \hat{K}_{t} \left( y_{\mbox{obs}} - \tilde{y}^{j}_{t} \right)$;

		Let $h^{j}_{t} = H(x^{j}_{t})$;
	}

	$\hat{\mu}^{x}_{t} = \frac{1}{M} \sum_{j=1}^M x^{j}_{t}$;
	
	$\hat{\mu}^{h}_{t} = \frac{1}{M} \sum_{j=1}^M h^{j}_{t}$;

	$\hat{C}^{x h}_{t} = \frac{1}{M-1} \sum_{j=1}^M \left( x^{j}_{t} - \hat{\mu}^{x}_{t} \right) \left( h^{j}_{t} - \hat{\mu}^{h}_{t} \right)^T$;

	$\hat{C}^{h h}_{t} = \frac{1}{M-1} \sum_{j=1}^M \left( h^{j}_{t} - \hat{\mu}^{h}_{t} \right) \left( h^{j}_{t} - \hat{\mu}^{h}_{t} \right)^T$;
}
\end{algorithm}

\subsection{Marginal likelihood estimators\label{subsec:Marginal-likelihood-estimators}}

In this section we describe three estimators for the normalising constant
of $\tilde{\pi}_{T}$:
\begin{equation}
Z_{T}=\int_{x}p(x)l(x)dx.\label{eq:enki_z}
\end{equation}

\subsubsection{Direct estimator}

The \emph{direct} estimator is derived from considering the ratio
of normalising constants in the recursive algorithm in the linear-Gaussian
case. We begin by noting that the incremental likelihood is given
by $l^{\alpha_{t}-\alpha_{t-1}}(x)=c_{t}g\left(y_{\text{obs}}\mid x,\gamma_{t}\Sigma_{y}\right)$,
where 
\[
c_{t}=\gamma_{t}^{d_{y}/2}\left(2\pi\right)^{\left(1-\gamma_{t}\right)d_{y}/2}\det\left(\Sigma_{y}\right)^{\left(1-\gamma_{t}\right)/2},
\]
given by the ratio of the normalising constants of the Gaussian to
the power $1/\gamma_{t}$ and the Gaussian with covariance $\gamma_{t}\Sigma_{y}$.
Then

\begin{eqnarray*}
\frac{Z_{t}}{Z_{t-1}} & = & \frac{\int_{x}p(x)l^{\alpha_{t}}(x)dx}{Z_{t-1}}\\
 & = & \int_{x}\frac{p(x)l^{\alpha_{t-1}}(x)}{Z_{t-1}}l^{\alpha_{t}-\alpha_{t-1}}(x)dx\\
 & = & c_{t}\int_{x}\pi_{t-1}(x)g\left(y_{\text{obs}}\mid x,\gamma_{t}\Sigma_{y}\right)dx\\
 & = & c_{t}\mathcal{N}\left(y_{\text{obs}}\mid\mu_{t-1}^{h},C_{t-1}^{hh}+\gamma_{t}\Sigma_{y}\right),
\end{eqnarray*}
using in the last step that the integral is the marginal distribution
of a multivariate Gaussian evaluated at $y_{\text{obs}}$. The direct
estimator then defined by using 
\begin{equation}
\hat{Z}_{T}^{d}=\prod_{t=1}^{T}\widehat{\frac{Z_{t}}{Z_{t-1}}}\label{eq:direct}
\end{equation}
with
\begin{equation}
\widehat{\frac{Z_{t}}{Z_{t-1}}}=c_{t}\mathcal{N}\left(y_{\text{obs}}\mid\hat{\mu}_{t-1}^{h},\hat{C}_{t-1}^{hh}+\gamma_{t}\Sigma_{y}\right).\label{eq:direct_ratio}
\end{equation}
In the linear/Gaussian case, convergence as $M\rightarrow\infty$
of ensemble Kalman techniques \citep{mandel_convergence_2011} yields
convergence of $\widehat{\frac{Z_{t}}{Z_{t-1}}}$ and hence $\hat{Z}_{T}$.
Note that to compute this estimate we need only run the loop in algorithm
\ref{alg:senki} up to iteration $t=T-1$.

\subsubsection{Unbiased estimator\label{subsec:Unbiased-estimator}}

The direct estimator is consistent, but not unbiased, since $\mathcal{N}\left(y_{\text{obs}}\mid\hat{\mu}_{t-1}^{h},\hat{C}_{t-1}^{hh}+\gamma_{t}\Sigma_{y}\right)$
is not an unbiased estimator of $\mathcal{N}\left(y_{\text{obs}}\mid\hat{\mu}_{t-1}^{h},\hat{C}_{t-1}^{hh}+\gamma_{t}\Sigma_{y}\right)$.
We can follow \citet{Price2017,drovandi_ensemble_2019} in devising
an unbiased version by instead using the estimator of \citet{ghurye_unbiased_1969},
which, for $\ensuremath{M>d_{y}+3}$ takes the form
\begin{eqnarray*}
\overline{\mathcal{N}}(y;\hat{\mu},\hat{\Sigma}) & = & (2\pi)^{-d_{y}/2}\frac{\rho(d_{y},M-2)}{\rho(d,M-1)(1-1/N)^{d_{y}/2}}|\left(M-1\right)\hat{\Sigma}|^{-(M-d_{y}-2)/2}\\
 &  & \qquad\psi\left(\left(M-1\right)\hat{\Sigma}-(y-\hat{\mu})(y-\hat{\mu})^{\top}/(1-1/M)\right)^{(M-d_{y}-3)/2},
\end{eqnarray*}
where
\[
\rho(k,v)=\frac{2^{-kv/2}\pi^{-k(k-1)/4}}{\prod_{i=1}^{k}\Gamma\left(\frac{1}{2}(v-i+1)\right)}
\]
and for a square matrix $A$, $\psi(A)=|A|$ if $A$ is positive definite
and $\psi(A)=0$ otherwise. The \emph{unbiased} estimator is then
given by

\begin{equation}
\hat{Z}_{T}^{u}=\prod_{t=1}^{T}\overline{\frac{Z_{t}}{Z_{t-1}}}\qquad\overline{\frac{Z_{t}}{Z_{t-1}}}=\prod_{t=1}^{T}c_{t}\overline{\mathcal{N}}\left(y_{\text{obs}}\mid\hat{\mu}_{t-1}^{h},\hat{C}_{t-1}^{hh}+\gamma_{t}\Sigma_{y}\right),\label{eq:unbiased_ratio}
\end{equation}
where the unbiasedness of this estimator holds only in the linear/Gaussian
case. We found empirically that, analogous to the finding in \citet{Price2017},
this estimator has similar properties to the direct estimator thus
we do not consider it further in this paper.

\subsubsection{Path sampling estimator}

\citet{Zhou2015} show how a path sampling estimator may be constructed
from SMC output; here we show that this is also possible for IEnKI.
\citet{Gelman1998a,Friel2008b} show that, for the path, over $\alpha\in[0,1]$,
of distributions proportional to $p(x)l^{\alpha}(x)$, under mild
regularity conditions
\[
\log\left(Z\right)=\int_{\alpha}\mathbb{E}_{\alpha}\left[\log l(.)\right]d\alpha,
\]
where $\mathbb{E}_{\alpha}$ denotes the expectation under the distribution
proportional to $p(x)l^{\alpha}(x)$, and $Z=\int_{x}p(x)l(x)dx$.
The sequence of distributions defined in equation (\ref{eq:annealed_target})
is a discretised form of this path. \citet{Zhou2015} propose to use
the points generated from an SMC algorithm on this sequence of targets
in to approximate $\log\left(Z\right)$ using a trapezoidal scheme
to estimate the integral. Here we propose the analogous idea for IEnKI
output:
\begin{equation}
\widehat{\log\left(Z_{T}\right)}^{p}=\sum_{t=1}^{T}\frac{1}{2\gamma_{t}}\left(U_{t}+U_{t-1}\right),\label{eq:path_sampling_estimator}
\end{equation}
where
\begin{eqnarray*}
U_{t} & = & \frac{1}{M}\sum_{j=1}^{M}\log l\left(x_{t}^{j}\right)\\
 & = & \frac{1}{M}\sum_{j=1}^{M}\log\left[\mathcal{N}\left(y_{\text{obs}}\mid h_{t}^{j},\Sigma_{y}\right)\right].
\end{eqnarray*}

\subsection{Alternative shifters}

The ``shift'' step in algorithm \ref{alg:senki} involves simulating
variables $\tilde{y}_{t}^{j}$ in order that the measurement noise
is correctly taken into account in the updates of the ensemble. Ensemble
Kalman filters are typically used in settings where the size of the
ensemble is small, thus the variance introduced by these simulations
can lead to large errors in estimates from the filter. As a remedy,
deterministic approaches were introduced: Monte Carlo implementations
of square-root Kalman filters \citep{tippett_ensemble_2003} that
ensure that the ensemble has consistent first and second moments.
In this paper we use both the \emph{square root} and \emph{adjustment}
shifters. Both take the form
\[
\ensuremath{x_{t}^{j}=x_{t-1}^{j}+\hat{K}_{t}\left(y_{\mbox{obs}}-\hat{\mu}_{t-1}^{h}\right)+b_{t}},
\]
where $b_{t}$ is chosen such that the covariance of the sample estimates
the desired posterior covariance \citep{whitaker_ensemble_2002-1,livings_unbiased_2008}.
The square root and adjustment filters produce produce different ensembles
that have the same covariance, thus exhibit different performance
when applied to non-linear/non-Gaussian models. We will refer to IEnKI
with each shifter as sIEnKI (stochastic), rIEnKI (square root) and
aIEnKI (adjustment). The rIEnK and aIEnKI algorithms are given the
Appendix.

\section{IEnKI-ABC\label{sec:IEnKI-ABC}}

In this section we investigate the use of iterative ensemble Kalman
inversion in ABC (IEnKI-ABC), specifically as an alternative estimator
of the ABC likelihood in equation \ref{eq:abc_llhd}. In section \ref{subsec:From-rare-event-ABC}
we introduce the new approach as an alternative to the SMC-based method
introduced for rare-event ABC in \citet{Prangle2016}, then in section
\ref{subsec:Sequences-of-targets} we describe an approach to automatically
chose the sequence of targets used in the method.

\subsection{Method\label{subsec:From-rare-event-ABC}}

\subsubsection{From rare-event ABC to IEnKI-ABC}

To estimate the ABC likelihood for parameter $\theta$ with tolerance
$\varepsilon$, the rare-event ABC approaches of \citet{Prangle2016}
and \citet{kerama_rare_2022} make use of the parameterisation in
equation (\ref{eq:re_abc_llhd}). \citet{Prangle2016} introduces
the idea of using an SMC algorithm on the sequence of targets $\tilde{\pi}_{t}(u\mid\theta)=K_{\varepsilon_{t}}\left(s_{\text{obs}}\mid S(G(\theta,u))\right)\phi(u\mid\theta)$
for $t=0:T$, to estimate the ABC likelihood at $\theta$, which is
equivalent to the normalising constant of $\tilde{\pi}_{T}$. Our
new approach uses the same sequence of distributions under a reparameterisation,
taking the unnormalised conditional (on $\theta$) posterior at iteration
$t$ to be
\begin{equation}
\tilde{\pi}_{t}(s\mid\theta)=f_{s}(s\mid\theta)K_{\varepsilon_{t}}\left(s_{\text{obs}}\mid s\right),\label{eq:ss_seq}
\end{equation}
where $f_{s}$ is the distribution of the transformed variable $S(x)$
with $x\sim f\left(\cdot\mid\theta\right)$. IEnKI-ABC then uses the
estimators in section \ref{sec:Iterative-ensemble-Kalman} to estimate
the integral
\begin{equation}
\int_{s}f_{s}(s\mid\theta)K_{\varepsilon}\left(s_{\text{obs}}\mid s\right)ds,\label{eq:conditional_abc_s}
\end{equation}
through the use of a sequence of distributions $\tilde{\pi}_{t}\left(s\mid\theta\right)=f_{s}(s\mid\theta)K_{\varepsilon_{t}}\left(s_{\text{obs}}\mid s\right)$
for $\infty=\varepsilon_{0}>...>\varepsilon_{T}=\varepsilon$. It is straightforward
to see that, for $\varepsilon>0$, this corresponds to a sequence of
temperatures in IEnKI of $\alpha_{t}=\left(\varepsilon/\varepsilon_{t}\right)^{2}$,
so that $\gamma_{t}=\varepsilon^{-2}\left(\varepsilon_{t}^{-2}-\varepsilon_{t-1}^{-2}\right)^{-1}$.
As an example of the use of these methods in this context, we show
IEnKI with a stochastic shifter in algorithm \ref{alg:senki-abc},
the output of which may be used for any of the estimators in section
\ref{subsec:Marginal-likelihood-estimators}.

\begin{algorithm} \caption{Stochastic iterative ensemble Kalman inversion for ABC, for a given $\theta$ and sequence $(\varepsilon_t)$.}\label{alg:senki-abc}

Simulate $M$ points, $\left\{ x_{0}^{j} \right\}_{j=1}^{M} \sim f(\cdot \mid \theta)$;\

For $j=1:M$, let $s^{j}_{0} = S(x^{j}_{0})$;

$\hat{\mu}^{s}_{0} = \frac{1}{M} \sum_{j=1}^M s^{j}_{0}$;

$\hat{C}^{s s}_{0} = \frac{1}{M-1} \sum_{j=1}^M \left( s^{j}_{0} - \hat{\mu}^{s}_{0} \right) \left( s^{j}_{0} - \hat{\mu}^{s}_{0} \right)^T$;

\For {$t=1:T$}
{
	$\hat{K}_{t} = \hat{C}^{s s}_{t-1} \left( \hat{C}^{s s}_{t-1} + \varepsilon^{-2}\left(\varepsilon_{t}^{-2}-\varepsilon_{t-1}^{-2}\right)^{-1} \Sigma_s \right)^{-1}$;

	\For{$n=1:N_a$}
	{
		Simulate $\tilde{s}^{j}_{t} \sim \mathcal{N}\left( s^{j}_{t-1}, \varepsilon^{-2}\left(\varepsilon_{t}^{-2}-\varepsilon_{t-1}^{-2}\right)^{-1} \Sigma_s \right)$;

		Shift $s^{j}_{t} = s^{(j)}_{t-1} + \hat{K}_{t} \left( s_{\mbox{obs}} - \tilde{s}^{j}_{t} \right)$;

	}

	$\hat{\mu}^{s}_{t} = \frac{1}{M} \sum_{j=1}^M s^{j}_{t}$;

	$\hat{C}^{s s}_{t} = \frac{1}{M-1} \sum_{j=1}^M \left( s^{j}_{t} - \hat{\mu}^{s}_{t} \right) \left( s^{j}_{t} - \hat{\mu}^{s}_{t} \right)^T$;
}
\end{algorithm}

\subsubsection{Remarks}

This approach has the following desirable properties:
\begin{itemize}
\item In contrast to rare-event ABC, IEnKI-ABC estimates the likelihood
through exploring the space of $s$, rather than $x$, which is considerably
easier given that we typically expect $d_{s}\ll d_{x}$.
\item The use of the shifting operations of IEnKI means that no proposals
for exploring $s$-space must be designed.
\item We observe that the use of IEnKI within ABC is more straightforward
than the general case presented in section \ref{sec:Iterative-ensemble-Kalman},
since the ``state'' ($x$) and ``measurement'' ($y$) spaces are
the same: no $H$ function is required to map $x$ to $y$. Non-linearity
in this function is one of the potential sources of error in estimators
based on IEnKI, but this issue is avoided in the ABC context.
\item Given simulations from $f$, the approach is very easy to implement.
The main tuning parameter is the sequence $(\varepsilon_{t})$, the choice
of which we describe in section \ref{subsec:Sequences-of-targets}.
\item Given that models commonly encountered in ABC are expensive to simulate,
the computational cost of the approach will often be dominated by
their simulation. If $\tau$ is the cost of simulating from $f$,
the complexity of: standard ABC is $O(M\tau)$; stochastic IEnKI-ABC
is $O(M\tau+Md_{s}^{2}+d_{s}^{3})$; square root IEnKI-ABC is $O(M\tau+Md_{s}^{2})$;
and adjustment IEnKI-ABC is $O(M\tau+M^{2}d_{s}+M^{3})$ \citep{whitaker_ensemble_2002-1}.
The scaling of adjustment IEnKI-ABC with $M$ means that it is only
of practical use in cases where $d_{s}<M$, which are not considered
in this paper.
\end{itemize}

\subsubsection{Comparison to alternative methods}

IEnKI has previously been used in the context of SBI by \citet{duffield_ensemble_2022},
where IEnKI was used on the parameter space $\theta$, rather than
in the ABC likelihood estimator. The main difference between our new
approach and this previous work is that \citet{duffield_ensemble_2022}
provides a Monte Carlo approximation of the true posterior only when
the joint distribution of $\theta$ and $s$ is Gaussian, whereas
IEnKI-ABC relies only on the conditional distribution $P_{s}$ of
$s\mid\theta$ being Gaussian.

Synthetic likelihood (SL) is an alternative to ABC introduced by \citet{Wood2010f}
and further developed by \citet{Price2017}. This approach also assumes
$P_{s}$ to be Gaussian, taking its mean and variance to be the sample
mean $\hat{\mu}^{s}$ and covariance $\hat{C}^{ss}$ from simulations
from $P_{s}$, and taking the likelihood to be $\mathcal{N}\left(s_{\text{obs}}\mid\hat{\mu}^{s},\hat{C}^{ss}\right)$
(or using the unbiased estimator in section \ref{subsec:Unbiased-estimator},
proposed in \citet{Price2017}).

SL is a limiting case of IEnKI-ABC, where the sequence of distributions
contains only $\pi_{0}$ and $\pi_{1}$, with $\varepsilon_{0}=\infty$
and $\varepsilon_{1}=0$. Here $K_{\varepsilon_{1}}$ is a Dirac delta and
the integral in equation \ref{eq:conditional_abc_s} collapses to
$f_{s}(s_{\text{obs}}\mid\theta)$. SL corresponds to implementing
lines 1-4 of algorithm \ref{alg:senki-abc} and using the result to
calculate the SL estimator. As we will see in section \ref{subsec:Gaussian-marginal-likelihood},
SL gives almost identical results to IEnKI-ABC when $P_{s}$ is Gaussian.
IEnKI-ABC is likely to improve upon the performance of SL when $P_{s}$
is non-Gaussian by using a sequence of distributions as described
in the paper. We observe such an improvement in section \ref{subsec:Lotka-Volterra-model}.

\subsection{Sequences of targets\label{subsec:Sequences-of-targets}}

\subsubsection{An adaptive procedure}

The choice of the temperatures $(\alpha_{t})$ signficantly affects
the performance of the approach. \citet{iglesias_bayesian_2018} introduces
the idea of adapting the sequence of temperatures as the algorithm
is running, following the corresponding idea in SMC introduced by
\citet{DelMoral2012g}. At the beginning of each iteration of the
algorithm, a new temperature $\alpha_{t}$ is found by using a bisection
algorithm that aims to ensure that the distance between successive
targets is not too large. In SMC, the \emph{effective sample size}
(ESS), calculated from particle weights, is used as an approximation
to the chi-square distance between successive targets. \citet{iglesias_bayesian_2018}
proposes to use the same approach, where weights assigned to ensemble
members (to be used only in this adaptive procedure) are defined to
be precisely those that would be assigned to particles in an SMC sampler
with MCMC moves. Ensemble member $j$ at iteration $t$ is assigned
(unnormalised) weight
\[
\tilde{w}_{t}^{j}=\exp\left(-\frac{\varepsilon_{t}^{-2}-\varepsilon_{t-1}^{-2}}{2}\left(s_{\text{obs}}-s_{t-1}^{j}\right)^{T}\Sigma_{s}^{-1}\left(s_{\text{obs}}-s_{t-1}^{j}\right)\right).
\]
Each weight is a function of the choice of $\varepsilon_{t}$, and is
calculated at each $\varepsilon_{t}$ considered in the bisection procedure.
The bisection aims to find $\varepsilon_{t}$ such that the ESS:
\[
\text{ESS}=\frac{\left(\frac{1}{M}\sum_{j=1}^{M}\tilde{w}_{t}^{j}\right)^{2}}{\frac{1}{M}\sum_{j=1}^{M}\left(\tilde{w}_{t}^{j}\right)^{2}}
\]
is a proportion $\beta$ of the sample size $M$.

As in SMC this adaptive procedure introduces a small bias into estimates
produced from the resultant IEnKI: the properties of the analogous
scheme for SMC are investigated empirically in \citet{Prangle2016}.
The approach has the appeal that it reduces the tuning of the sequence
of temperatures to the choice of a single parameter $\beta$. However,
the length of the sequence chosen for parameters with a small likelihood
can be large (resulting in additional computation), so we instead
consider an alternative approach, tailored to the particular sequence
of targets in equation (\ref{eq:ss_seq}).

\subsubsection{Choice of tempering scheme\label{subsec:Choice-of-tempering}}

Consider an infinite sequence of (unnormalised) distributions $\tilde{\pi}_{t}(s\mid\theta)=P_{s}(s\mid\theta)K_{\varepsilon_{t}}\left(s_{\text{obs}}\mid s\right)$
where $\varepsilon_{t}\rightarrow0$, recalling that $K_{\varepsilon_{t}}\left(s_{\text{obs}}\mid s\right)$
is Gaussian with mean $s$ and covariance $\varepsilon_{t}^{2}\Sigma_{s}$.
For a fixed $\theta$, this sequence has similar properties to a sequence
of Bayesian posterior distributions with an increasing number $n$
of data points, with $P_{s}\left(\cdot\mid\theta\right)$ playing
the role of the prior and $K_{\varepsilon_{t}}$ the role of a Gaussian
likelihood with mean $s_{\text{obs}}$, covariance $\Sigma_{s}$ with
$n=\varepsilon_{t}^{-2}$ observations. As long as $P_{s}(\cdot\mid\theta)$
is absolutely continuous in a neighbourhood of $s_{\text{obs}}$ with
continuous positive density at $s_{\text{obs}}$, direct application
of the Bernstein-von Mises theorem (\citet{vaart_asymptotic_1998};
Chapter 10.2) yields that as $t\rightarrow\infty$ (i.e. $\varepsilon_{t}\rightarrow0$
equating to $n\rightarrow\infty$), $\tilde{\pi}_{t}(s\mid\theta)\rightarrow\mathcal{\mathcal{MVN}}\left(s_{\text{obs}},\varepsilon_{t}^{2}\Sigma_{s}\right)$.

In this section we make use of this property, together with the recent
work of \citet{chopin_connection_2024} which suggests a principled
approach to choosing a tempering sequence $(\alpha_{t})$, and hence
a sequence $(\varepsilon_{t})$, through connection to mirror descent.
We follow this approach, using the assumption that the sequence of
targets is Gaussian. We suppose that $\pi_{0}$ and $\pi_{T}$ are
both multivariate normal distributions with common mean $s_{\text{obs}}$
and covariance $\Sigma_{0}$ and $\varepsilon^{2}\Sigma_{s}$ respectively.
Through the argument above, the normality assumption for $\pi_{T}$
is likely to be realistic when $\varepsilon$ is small. Recall that $\Sigma_{s}$
is a diagonal matrix chosen to scale the summary statistics. We assume
that $\Sigma_{0}$ is some multiple $\kappa^{2}$ of $\Sigma_{s}$:
this is not unreasonable, since in practice a common choice of $\Sigma_{s}$
is to use the diagonal entries of an estimate $\hat{\Sigma}_{0}$,
obtained through simulation from $f$, of $\Sigma_{0}$ (in this case
we would have $\kappa^{2}=1$). In the empirical studies below, we
take $\kappa$ to be the mean of $\hat{\sigma}_{i,0}/\sigma_{i,s}$,
where $\sigma_{i,s}$ is the square root of the $i$th element on
the diagonal of $\Sigma_{s}$ and $\hat{\sigma}_{i,0}$ is the sample
standard deviation of the $i$th statistic over the $M$ simulations
from $P_{s}$. The argument of \citet{chopin_connection_2024} (details
in the Appendix) results in continuous tempering schedule of
\begin{equation}
\alpha(t)=\exp\left(2\log\left(\frac{\kappa}{\varepsilon}\right)t+\log\left(\frac{\varepsilon^{2}}{\kappa^{2}-\varepsilon^{2}}\right)\right)-\frac{\varepsilon^{2}}{\kappa^{2}-\varepsilon^{2}}\label{eq:sequence}
\end{equation}
 for $t\in[0,1]$. We propose using a discretised version $(\alpha_{t})$
of this schedule, taking $\alpha_{t}=\alpha(t/T)$ for $t=0:T$.

\subsubsection{Target skipping\label{subsec:Target-skipping}}

The result that $\tilde{\pi}_{t}(s\mid\theta)\rightarrow\mathcal{\mathcal{MVN}}\left(s_{\text{obs}},\varepsilon_{t}^{2}\Sigma_{s}\right)$
has a further implication: when using a finite sequence of targets
$t=0:T$ in IEnKI, when $\varepsilon_{T}$ is chosen to be small, we
might expect that for some $t<T$ the ensemble representing a from
$\pi_{t}(s\mid\theta)$ will be close to a sample from a Gaussian
distribution. Recall that the motivation for using the iterative EnKI
algorithm arose purely from the acknowledgement that we do not expect
the sequence of targets to be Gaussian. As soon as we detect that
a target $\pi_{t}(s\mid\theta)$ is close to Gaussian, we may also
assume that targets for $t'>t$ are also close to Gaussian, and we
may use a single EnK update step to move our ensemble from $\pi_{t}(s\mid\theta)$
to $\pi_{T}(s\mid\theta)$. We envisage that this approach will lead
to computational savings without sacrifiing much accuracy.

Based on this principle, we propose the following approach. At each
iteration of algorithm \ref{alg:senki-abc}, directly after line 5,
we perform a hypothesis test where the null hypothesis is that the
ensemble is drawn from a multivariate Gaussian distribution. If the
null hypothesis is accepted at iteration $t$, in the remainder of
the IEnKI we ``skip'' targets $t+1:T-1$ and perform the update
(lines 6-10) moving directly to target $\pi_{T}$. In this paper we
used the Henze-Zirkler test \citep{henze_class_1990} which has been
illustrated to be powerful in a range of scenarios (e.g. \citep{farrell_tests_2007,ebner_tests_2020}.

\section{Applications\label{sec:Applications}}

\subsection{Gaussian marginal likelihood\label{subsec:Gaussian-marginal-likelihood}}

In this section we investigate the empirical properties of the new
marginal likelihood estimators introduced in this paper. We study
a toy example where the true marginal likelihood is available, and
where the Gaussian assumptions underpinning IEnKI are all satisfied.
We take $P_{s}(s\mid\theta)=\mathcal{N}(s\mid\theta,1)$, $K_{\varepsilon}(s_{\text{obs}}\mid s)=\mathcal{N}(s_{\text{obs}}\mid s,\varepsilon^{2})$,
so that the exact ABC likelihood is available via
\[
\int_{s}K_{\varepsilon}\left(s_{\text{obs}}\mid s\right)P_{s}\left(s\mid\theta\right)ds=\mathcal{N}\left(s_{\text{obs}}\mid\theta,1+\varepsilon^{2}\right).
\]
In this section we use $s_{\text{obs}}=0$ and $\theta=0$ and examine
the properties of ABC, SL and IEnKI-ABC estimates of the ABC likelihood.
To accurately compare SL with the other two approaches we set the
data generating process used in this approach to be $\mathcal{N}\left(\cdot\mid0,1+\varepsilon^{2}\right)$.
We compared the algorithms for tolerances $\varepsilon=0.1,0.01,0.001,0.0001$,
$M=10,50,100,200$ points simulated from $P_{s}$. and, for IEnKI-ABC,
$T=5,10,20$ targets (plus $T=50,100,200$ for the path sampling estimator).
The stochastic and square root shifting approaches were compared for
IEnKI-ABC, using the direct (equation (\ref{eq:direct})) and path
sampling (equation (\ref{eq:path_sampling_estimator})) estimators.
Since the model is linear/Gaussian, the square root and adjustment
approaches produce almost identical results, thus the adjustment approach
is omitted. All experiments were performed using the \href{https://github.com/richardgeveritt/ilike}{ilike}
R package, with the model written in C++. Each algorithm was run 100
times, and the bias, standard deviation, root mean squared error (RMSE)
of IEnKI marginal likelihood estimators was estimated from these 100
runs. In this section we examine the properties of the direct estimator
from IEnKI-ABC as $\varepsilon$ changes, and as $T$ changes.

\begin{figure}
\subfloat[$\log$ RMSE as a function of the $\log$ tolerance $\varepsilon$.\label{fig:RMSE-as-a-2}]{\includegraphics[scale=0.45]{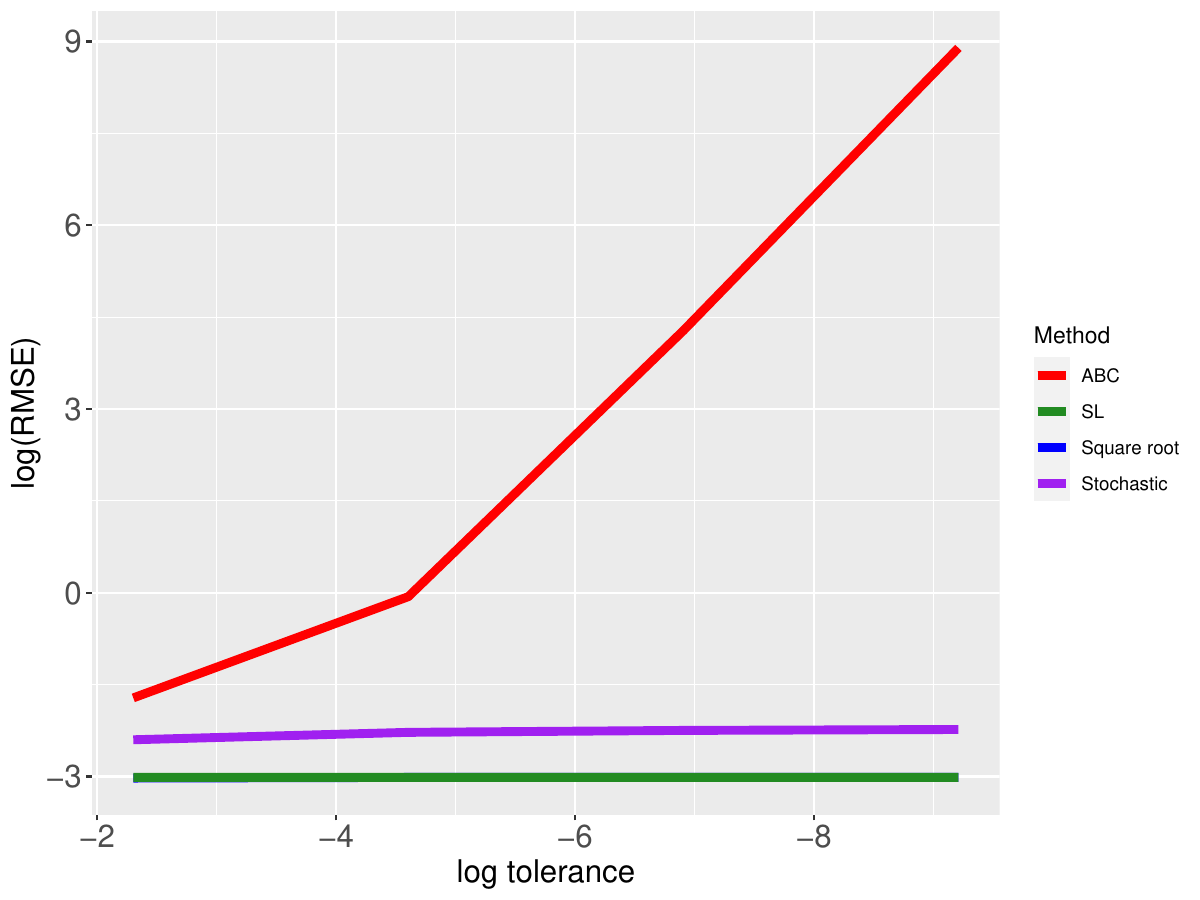}

}\subfloat[$\log$ RMSE as a function of the number of targets $T$.\label{fig:RMSE-as-a-1}]{\includegraphics[scale=0.45]{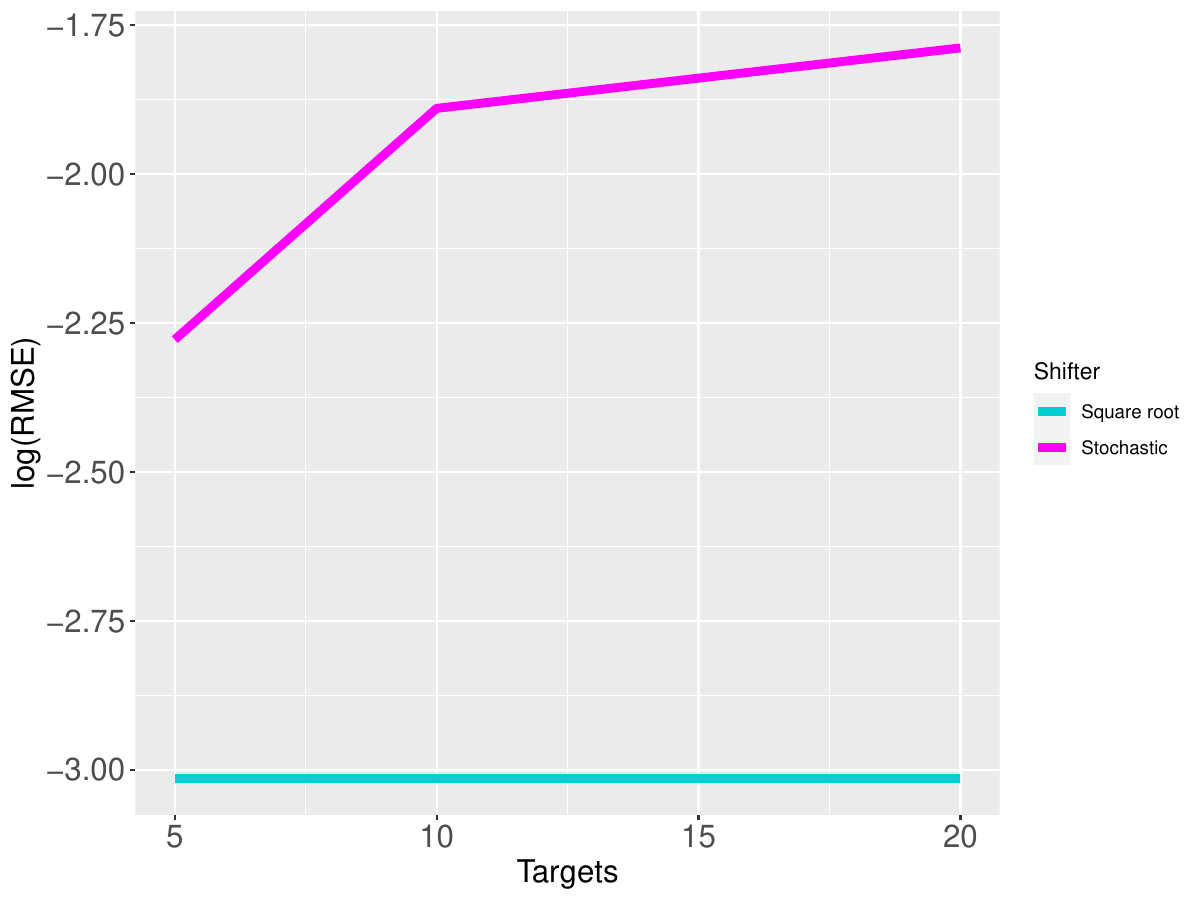}

}

\caption{Properties of IEnKI normalising constant estimators on the Gaussian
example.}
\end{figure}

Figure \ref{fig:RMSE-as-a-2} shows how the RMSE of the different
approaches scales with $\varepsilon$, where $M=200$ and $T=5$. We
observe a dramatic increase in RMSE for ABC as $\varepsilon$ gets closer
to 0; the other schemes have the desirable property that the RMSE
is similar for all values of $\varepsilon$. The rIEnKI-ABC and SL results
are identical up to rounding errors, whereas we observe additional
error, due to the additional variance introduced by the shifter in
sIEnKI-ABC. Figure \ref{fig:RMSE-as-a-1}, shows the $\log$ RMSE
of the direct estimator for $M=200$ and $\varepsilon=0.01$, clearly
showing the effect on the direct estimator from sIEnKI-ABC of increasing
the number of targets $T$. We observe an increased variance as $T$
increases for sIEnKI-ABC, whereas rIEnKI-ABC and aIEnKI-ABC are not
affected by changes in $T$. In this example, where a conjugate Gaussian
update holds, the best choice for $T$ (giving lower variance and
computational cost) is 1. The non-Gaussian example in section \ref{subsec:Lotka-Volterra-model}
will show a case where it is useful to choose $T>1$.

Further results and discussion are found in the Appendix, including
an illustration that the error of path sampling IEnKI estimators increases
as $\varepsilon$ decreases, reducing the appeal of this approach (compared
to the direct estimator) for use in ABC.

\subsection{Lotka-Volterra model\label{subsec:Lotka-Volterra-model}}

In this section we apply IEnKI-ABC to the well-studied stochastic
Lotka-Volterra (LV) predator-prey model. We use the Markov jump process
version of the model with three parameters, as previously studied
in \citet{Wilkinsona,Papamakarios2016}. Let $x_{1}$ be the number
of predators and $x_{2}$ the number of prey, evolving with the following
reactions:
\begin{itemize}
\item A prey may be born, with rate $\theta_{1}x_{2}$, increasing $x_{2}$
by one.
\item The predator-prey interaction in which $x_{1}$ increases by one and
$x_{2}$ decreases by one, with rate $\theta_{2}x_{1}x_{2}$.
\item A predator may die, with rate $\theta_{3}x_{1}$, decreasing $x_{1}$
by one.
\end{itemize}
ABC is a popular approach to inference in such models, even though
it can be significantly less efficient than other approaches, such
as particle MCMC. In this section we take the opportunity to compare
several different approaches, alongside IEnKI-ABC. Let $x_{t,1}$
and $x_{t,2}$ be, respectively, the number of predator and prey at
time $t$ and $x_{t}=\left(x_{t,1},x_{t,2}\right)^{T}$. We simulate
the LV model using the Gillespie algorithm via C++ code adapted from
the \texttt{smfsb} R package \citep{wilkinson_smfsb_2024}, using
the \texttt{LVperfect} data in the \texttt{smfsb} package to be our
observed data $y_{\text{obs}}$, consisting of measurements of $x_{t,1}$
and $x_{t,2}$ at each time $t$ (consisting of the the initial states
at time 0, plus measurements at 15 further times). We use the same
model for every inference algorithm we test, with the model having
slightly different interpretations for the different algorithms used
to estimate the marginal likelihood of $\theta$:
\begin{itemize}
\item For ABC and IEnKI-ABC we take $P$ to be the LV model so that $y=x$
(i.e. simulations from the LV model are treated as our observed data),
take $s_{\text{}}=y$ (i.e. use the full data rather than summary
statistics), and choose $K_{\varepsilon}$ to be Gaussian with covariance
$\varepsilon^{2}\Sigma_{s}$, where $\Sigma_{s}$ is taken to be the
identity.
\item For SL, we take $y_{t,1}\sim\mathcal{N}\left(\cdot\mid x_{t,1},\varepsilon^{2}\right)$,
$y_{t,2}\sim\mathcal{N}\left(\cdot\mid x_{t,2},\varepsilon^{2}\right)$
to be noisy measurements of $x_{t,1},x_{t,2}$, which are simulated
from the LV model for $t=0:15$. We again take $s=y$.
\item We also use a bootstrap particle filter (PF) \citep{Gordon1993} and
ensemble Kalman filters (EnKF) \citep{evensen_sequential_1994}. These
filters also take $y_{t,1}\sim\mathcal{N}\left(\cdot\mid x_{t,1},\varepsilon^{2}\right)$,
$y_{t,2}\sim\mathcal{N}\left(\cdot\mid x_{t,2},\varepsilon^{2}\right)$
to be noisy measurements of $x_{t,1},x_{t,2}$ simulated from the
LV model, and perform filtering on $x$ given $y$. We use EnKFs with
stochastic, square root and adjustment shifters - previously only
the stochastic shifter has been investigated in the context of marginal
likelihood estimation \citep{drovandi_ensemble_2019}.
\end{itemize}

\subsubsection{Marginal likelihood estimation}

In this section we investigate the empirical properties of the marginal
likelihood estimators mentioned above, in a similar manner to section
\ref{subsec:Gaussian-marginal-likelihood}. This section provides
a clearer picture of the relative efficiency of IEnKI-ABC in a more
realistic situation, using a model that does not satisfy the Gaussian
assumptions outlined in section \ref{subsec:Iterative-ensemble-Kalman}.
All experiments were performed using the \href{https://github.com/richardgeveritt/ilike}{ilike}
R package, with the model written in C++. In this section we study
the standard deviation (SD) of the likelihood estimator, since this
is the main determining factor in the efficiency of an MCMC algorithm
using the likelihood estimator, and we found the bias is difficult
to estimate accurately.

We estimated the likelihood at the point at which the \texttt{LVperfect}
data was generated, $\theta=(1,0.005,0.6)^{T}$. All methods used
$M=100$ points: for the PF and EnKF this corresponds to the number
of particles and ensemble members respectively. We used EnKF with
stochastic (sEnKF), square root (rEnKF) and adjustment (aEnKF) shifters.
We also compared three IEnKI-ABC estimators, all using the stochastic
shifter and with $T=100$: the direct estimator (labelled ``sIEnKI-ABC'');
the path estimator (``sIEnKI-ABCpath'') and the direct estimator
where the target skipping approach is used with significance $\alpha=0.1$
(``sIEnKI-ABCskip''). Results for the deterministic shifters in
IEnKI-ABC are not presented here: poorly conditioned covariance matricies
(due to diverging LV simulations) rendered these methods inaccurate.
The cost of most of the methods was dominated by the cost of simulating
from the LV model $M=100$ times, taking approximately 0.1s. The cost
of aEnKF was marginally higher at 0.13s, and the use of $T=100$ targets
increased the cost of the IEnKI methods, with sIEnKI-ABC taking 0.25s
and sIEnKI-ABCpath 0.34s, except when using sIEnKI-ABCskip whose cost
was again 0.1s. \href{https://youtu.be/nERovsRTDzI}{This video} illustrates
a run of sIEnKI-ABC for in this situation.

\begin{figure}
\subfloat[$\log$ SD as a function of the $\log$ tolerance $\varepsilon$.\label{fig:RMSE-as-a-3}]{\includegraphics[scale=0.45]{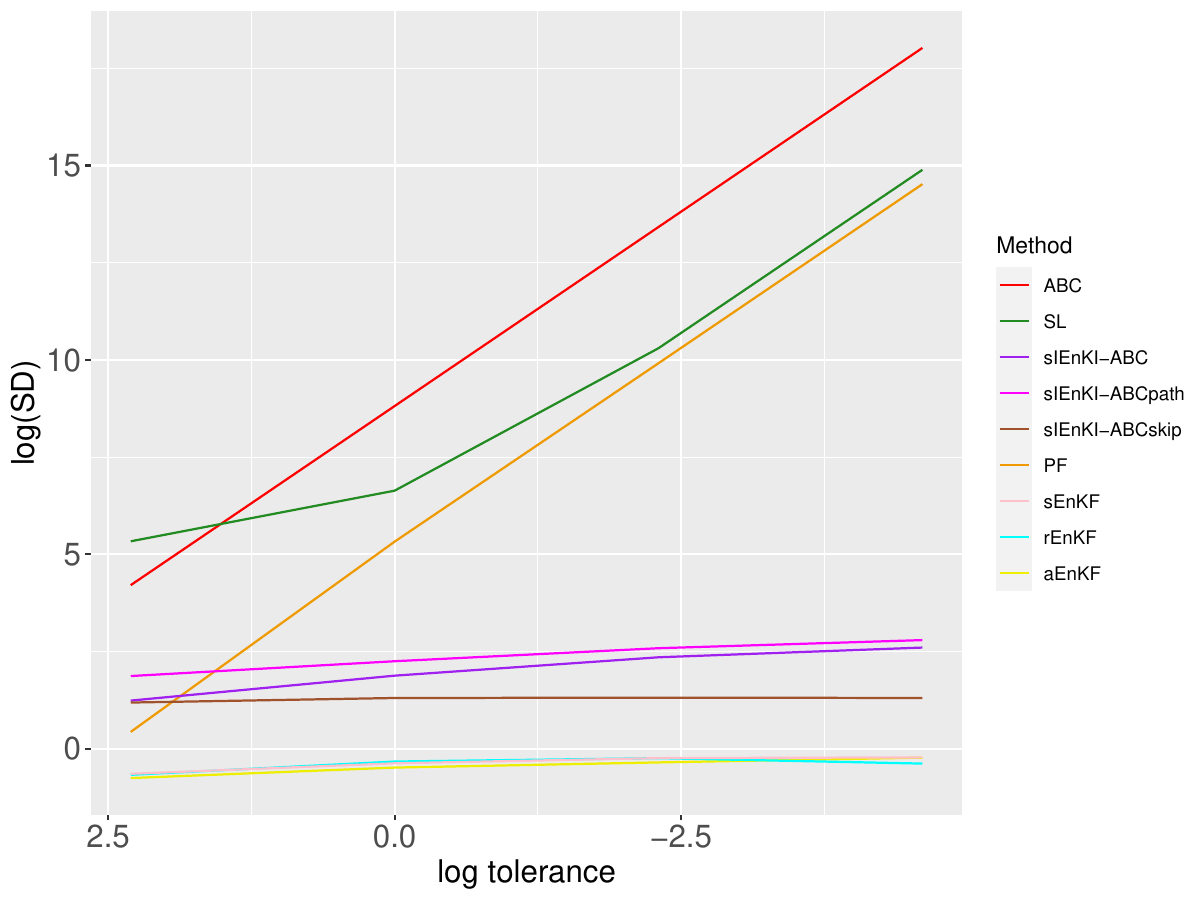}

}\subfloat[$\log$ SD of estimator as a function of the number of targets $T$.\label{fig:RMSE-as-a-1-1}]{\includegraphics[scale=0.45]{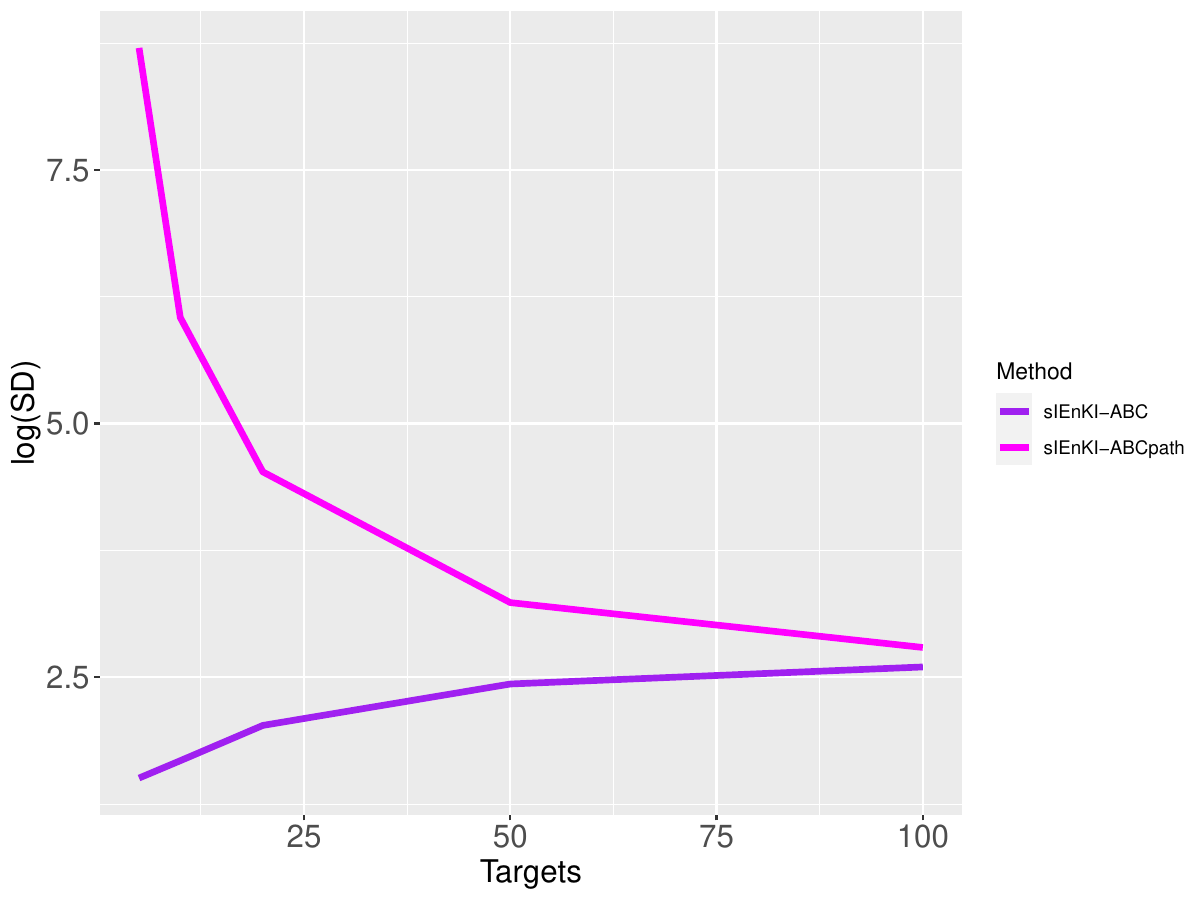}

}

\caption{Properties of alternative likelihood estimators on the Lotka-Volterra
example.}
\end{figure}

In figure \ref{fig:RMSE-as-a-3} we compare the $\log$ SD of the
likelihood estimators from each of the methods mentioned in the previous
section, as a function of the $\log$ of $\varepsilon$. ABC and PF exhibit
apparently exponentially increasing SD with the decrease in $\varepsilon$,
with SL also following this trend when $\varepsilon$ is small. The EnKF
exhibits the best performance for all $\varepsilon$ and the SD only
slightly increases as $\varepsilon$ decreases. The deterministic shifters
offer slight improvements over the stochastic shifter. The SD of the
sIEnKI-ABC approaches is also not dramatically affected by the decrease
in $\varepsilon$. The target skipping approach results in a smaller
SD due to omitting IEnKI steps that would not dramatically change
the likelihood estimates but would result in additional noise via
the stochastic shifter.

Figure \ref{fig:RMSE-as-a-1-1} examines the effect on sIEnKI-ABC
and sIEnKI-ABCpath of changing the number of targets $T$ (for $\varepsilon=0.1$
and $M=100$). For the path estimator we observe that a relatively
large number of targets are required in order for the numerical integration
approach to achieve a small error. For the direct approach there is
again an increase in the error as $T$ increases due to the stochastic
shifter, as observed in section \ref{subsec:Gaussian-marginal-likelihood}.
However, taking $T>1$ results in significant benefits compared to
the $T=1$ case, represented in figure \ref{fig:RMSE-as-a-3} by the
SL results which are very poor due to $f_{s}$ being far from Gaussian.

\subsubsection{Use in MCMC}

\begin{figure}
\subfloat[Estimated marginal posterior densities for the Lotka-Volterra model
when $\varepsilon=10$.]{\includegraphics[scale=0.45]{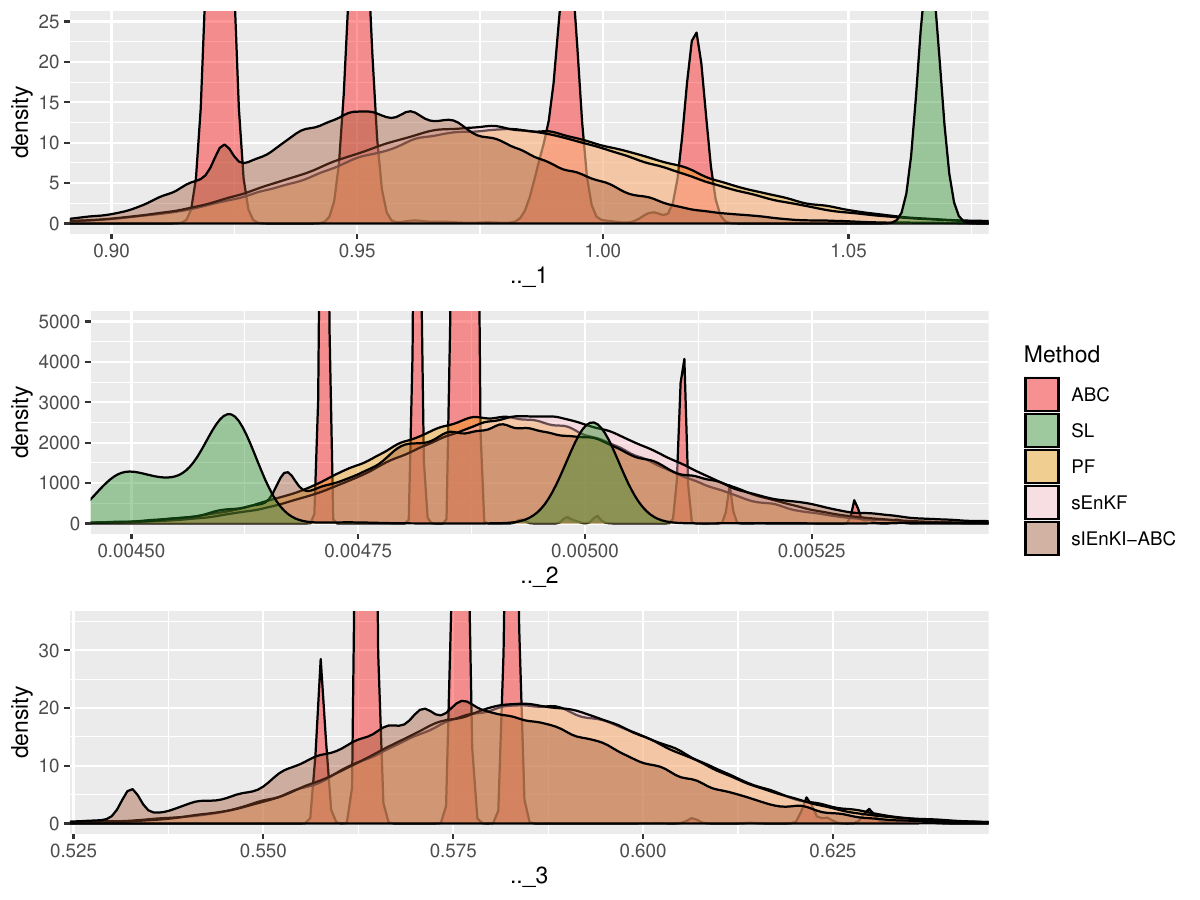}

}\subfloat[Estimated marginal posterior densities for the Lotka-Volterra model
when $\varepsilon=0.1$.]{\includegraphics[scale=0.45]{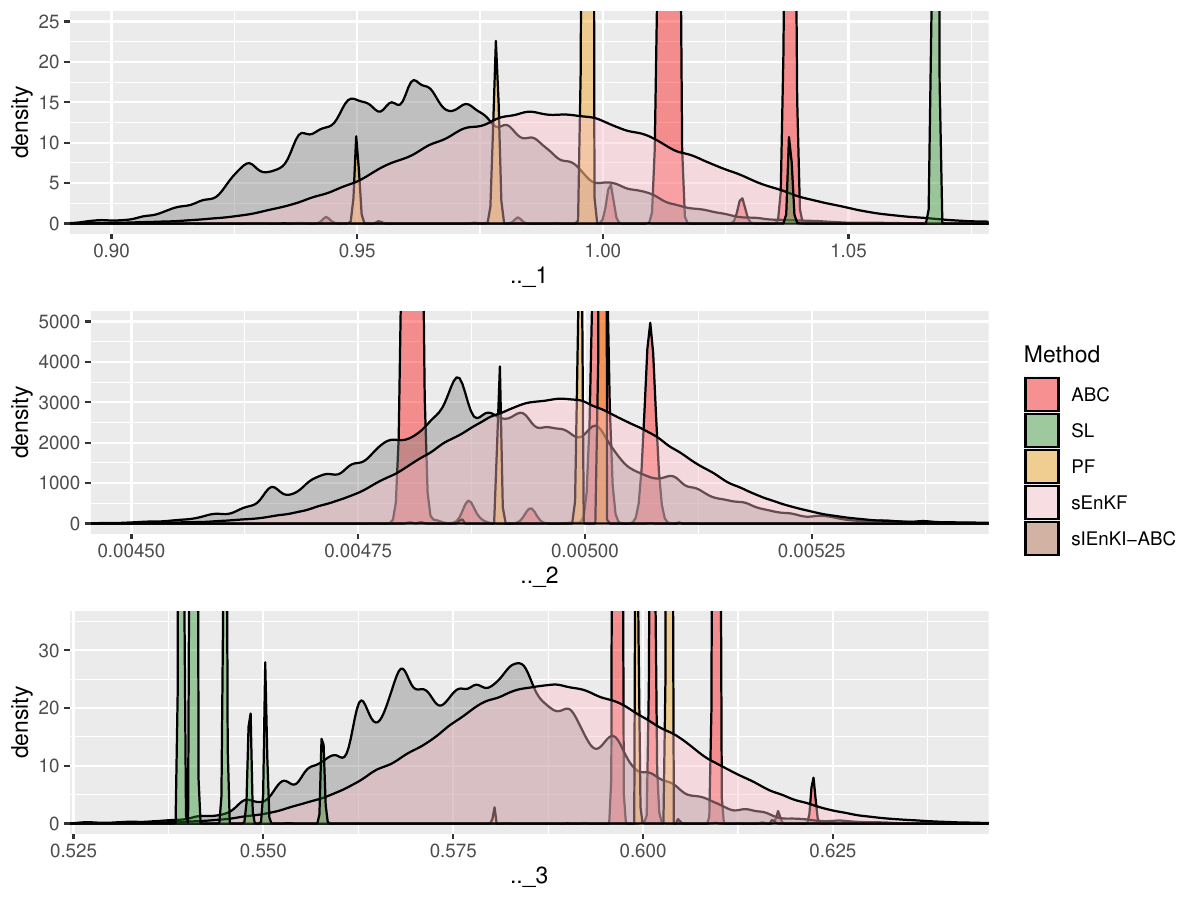}

}

\caption{Estimated marginal posterior densities for the Lotka-Volterra model
when $\varepsilon=10$ (left) and $\varepsilon=0.1$ (right).\label{fig:Estimated-marginal-posterior}}
\end{figure}

We ran Metropolis-Hastings (MH) algorithms using each of the likelihood
estimators in the previous section, taking $M=100$ for all algorithms.
Each algorithm was run for $10^{6}$ iterations and was initialised
at $\theta=(1,0.005,0.6)^{T}$. For the IEnKI-ABC approaches, we used
only the target skipping approach, taking $T=100$ and $\alpha=0.01$
which results in a comparable computational cost for all algorithms
used in this section. A multivariate Gaussian proposal was used, with
a covariance determined through pilot runs. The estimated posterior
distributions for $\varepsilon=10$ and $\varepsilon=0.1$ are shown in
figure \ref{fig:Estimated-marginal-posterior} and the multiESS (from
the \texttt{mcmcse} R package) for each run in table \ref{tab:The-multiESS-for}.
For several algorithms there are only a few acceptances in the whole
run: for these the multiESS cannot be estimated reliably and so we
report it simply as $<50$.

ABC-MCMC and SL-MCMC perform very poorly for both choices of $\varepsilon$.
Particle marginal MH (using the PF estimate in MH) performs well for
$\varepsilon=10$, but poorly for $\varepsilon=0.1$. Ensemble MCMC (using
the EnKF estimate in MH) performs very well for both choices of $\varepsilon$.
For $\varepsilon=0.1$ IEnKI-ABC-MCMC significantly outperforms all approaches
except ensemble MCMC, although some bias is evident in the IEnKI-ABC-MCMC
posteriors.

\begin{table}
\begin{tabular}{|c|c|c|c|c|c|}
\hline 
$\varepsilon$ & ABC & SL & PF & EnKF & sEnKI-ABC\tabularnewline
\hline 
\hline 
10 & <50 & <50 & 27307 & 56665 & 6085\tabularnewline
\hline 
0.1 & <50 & <50 & <50 & 47181 & 2408\tabularnewline
\hline 
\end{tabular}

\caption{The multiESS for each MCMC algorithm for the Lotka-Volterra model.\label{tab:The-multiESS-for}}
\end{table}

\section{Discussion\label{sec:Discussion}}

This paper introduces new, straightforward to implement, methods for
estimating the ABC likelihood based on IEnKI that we show empirically
to have substantially lower variance than the standard approach, in
particular we see that the variance grows much more slowly as $\varepsilon\rightarrow0$.
The direct IEnKI-ABC estimator is also related to the SL approach.
For cases where the summary statistic distribution $f_{s}$ is close
to Gaussian, the performance of IEnKI-ABC is likely to be similar
to that of SL, however we see empirically that IEnKI-ABC outperforms
SL for the LV model when $f_{s}$ is non-Gaussian. IEnKI-ABC uses
sequence of target distributions analogous to the sequence used in
SMC approaches. We introduce an approach to choosing this sequence
automatically, leaving the only tuning parameter to be the number
of targets, and also introduce a method for automatically skipping
to the end of the target sequence to save computation and reduce variance.
IEnKI-ABC is applicable in all cases where ABC may be applied, thus
is more generally applicable that some alternative approaches: for
example the EnKF methods used in section \ref{subsec:Lotka-Volterra-model}.

Since IEnKI is being used outside of the linear-Gaussian setting,
estimators based on its output will have a bias that is difficult
to quantify: a feature avoided when using rare-event ABC. In particular,
we might expect a large bias in situations where members of the sequence
$\tilde{\pi}_{t}\left(s\mid\theta\right)$ are not unimodal. However,
we expect in many applications that IEnKI-ABC posteriors are likely
to have lower bias than those from SL, which are more influenced by
the Gaussian assumption, and ABC, due to the necessity in using a
large tolerance.

\appendix

\section{IEnKI with deterministic shifters}

IEnKI algorithms using \emph{square root} and \emph{adjustment} filters
are given in algorithms \ref{alg:renki} and \ref{alg:aenki} respectively.

\begin{algorithm} \caption{Square root iterative ensemble Kalman inversion.}\label{alg:renki}

Simulate $M$ points, $\left\{ x_{0}^{j} \right\}_{j=1}^{M} \sim p$;\

For $j=1:M$, let $h^{j}_{0} = H(x^{j}_{0})$;

$\hat{\mu}^{x}_{0} = \frac{1}{M} \sum_{j=1}^M x^{j}_{0}$;
	
$\hat{\mu}^{h}_{0} = \frac{1}{M} \sum_{j=1}^M h^{j}_{0}$;

$\hat{C}^{x h}_{0} = \frac{1}{M-1} \sum_{j=1}^M \left( x^{j}_{0} - \hat{\mu}^{x}_{0} \right) \left( h^{j}_{0} - \hat{\mu}^{h}_{0} \right)^T$;

$\hat{C}^{h h}_{0} = \frac{1}{M-1} \sum_{j=1}^M \left( h^{j}_{0} - \hat{\mu}^{h}_{0} \right) \left( h^{j}_{0} - \hat{\mu}^{h}_{0} \right)^T$;

\For {$t=1:T$}
{
	$\hat{K}_{t} = \hat{C}^{x h}_{t-1} \left( \hat{C}^{h h}_{t-1} + \gamma_t \Sigma_y \right)^{-1}$;

	$\hat{S}_{t} = \hat{C}^{h h}_{t-1} + \gamma_t \Sigma_y$;

	$\bar{K}_{t} = \hat{C}^{x h}_{t-1} \left( \hat{S}^{1/2}_{t} + \sqrt{\gamma_t} \Sigma^{1/2}_y \right)^{-1}$;

	\For{$n=1:N_a$}
	{

		Shift $x^{j}_{t} = x_{t-1}^{j}+\hat{K}_{t}\left(y_{\mbox{obs}}-\hat{\mu}_{t-1}^{h}\right)-\bar{K}_{t}\left(h_{t-1}^{(j)}-\hat{\mu}_{t-1}^{h}\right)$;

		Let $h^{j}_{t} = H(x^{j}_{t})$;

	}

    $\hat{\mu}^{x}_{t} = \frac{1}{M} \sum_{j=1}^M x^{j}_{t}$;
	
	$\hat{\mu}^{h}_{t} = \frac{1}{M} \sum_{j=1}^M h^{j}_{t}$;

	$\hat{C}^{x h}_{t} = \frac{1}{M-1} \sum_{j=1}^M \left( x^{j}_{t} - \hat{\mu}^{x}_{t} \right) \left( h^{j}_{t} - \hat{\mu}^{h}_{t} \right)^T$;

	$\hat{C}^{h h}_{t} = \frac{1}{M-1} \sum_{j=1}^M \left( h^{j}_{t} - \hat{\mu}^{h}_{t} \right) \left( h^{j}_{t} - \hat{\mu}^{h}_{t} \right)^T$;

}
\end{algorithm}

\begin{algorithm} \caption{Adjustment iterative ensemble Kalman inversion.}\label{alg:aenki}

Simulate $M$ points, $\left\{ x_{0}^{j} \right\}_{j=1}^{M} \sim p$;\

For $j=1:M$, let $h^{j}_{0} = H(x^{j}_{0})$;

$\hat{\mu}^{x}_{0} = \frac{1}{M} \sum_{j=1}^M x^{j}_{0}$;

$\hat{\mu}^{h}_{0} = \frac{1}{M} \sum_{j=1}^M h^{j}_{0}$;

$\hat{Z}^{x}_{0} = \frac{1}{\sqrt{M-1}} \sum_{j=1}^M \left( x^{j}_{0} - \hat{\mu}^{x}_{0} \right)$;

$\hat{Z}^{h}_{0} = \frac{1}{\sqrt{M-1}} \sum_{j=1}^M \left( h^{j}_{0} - \hat{\mu}^{h}_{0} \right)$;

$P_{0} W_{0}^{1/2} V_{0}^T = \mbox{SVD}\left( \hat{Z}^{x}_{0} \right)$;

\For {$t=1:T$}
{

	$U^T_{t} D_{t} U_{t} = \mbox{SVD}\left( V_{t-1}^T \left( I_{d_x} + (\hat{Z}^{h}_{t-1})^T (\gamma_t \Sigma_y)^{-1} \hat{Z}^{h}_{t-1} \right) V_{t-1} \right)$;

	$\hat{K}_{t} =  \hat{Z}^{x}_{t-1} (\hat{Z}^{h}_{t-1})^T \left( \hat{Z}^{h}_{t-1} (\hat{Z}^{h}_{t-1})^T + \gamma_t \Sigma_y \right)^{-1}$;

	$A_t = P_{t-1}W_{t-1}^{1/2} U_t D_t^{1/2} W_{t-1}^{-1/2} P_{t-1}^T$;

	\For{$n=1:N_a$}
	{

		Shift $x^{j}_{t} = \hat{\mu}_{t-1}^{x} + \hat{K}_{t} \left( y_{\mbox{obs}} - \hat{\mu}^{h}_{t-1} \right) + A_t \left(x^{(j)}_{t-1}- \hat{\mu}^{x}_{t-1}\right)$;

		Let $h^{j}_{t} = H(x^{j}_{t})$;

	}

	$\hat{\mu}^{x}_{t} = \frac{1}{M} \sum_{j=1}^M x^{j}_{t}$;

	$\hat{\mu}^{h}_{t} = \frac{1}{M} \sum_{j=1}^M h^{j}_{t}$;

	$\hat{Z}^{x}_{t} = \frac{1}{\sqrt{M-1}} \sum_{j=1}^M \left( x^{j}_{t} - \hat{\mu}^{x}_{t} \right)$;

	$\hat{Z}^{h}_{t} = \frac{1}{\sqrt{M-1}} \sum_{j=1}^M \left( h^{j}_{t} - \hat{\mu}^{h}_{t} \right)$;

	$P_{t} W_{t}^{1/2} V_{t}^T = \mbox{SVD}\left( \hat{Z}^{x}_{t} \right)$;
}
\end{algorithm}

Each approach has a different computational complexity: $O(Md_{x}d_{y}+Md_{y}^{2}+d_{y}^{3}+d_{x}d_{y}^{2})$
for sIEnKI, $O(Md_{y}+Md_{x}d_{y})$ for rIEnKI and $O(M^{2}d_{y}+M^{3}+M^{2}d_{x})$
for aIEnKI \citep{whitaker_ensemble_2002-1}.

\section{Sequence of targets for IEnKI-ABC\label{sec:Sequence-of-targets}}

When $d_{s}$ is large, \citet{chopin_connection_2024} suggests deriving
a tempering sequence by solving the ODE
\begin{equation}
\dot{\alpha}=cI(\alpha)^{-1/2},\label{eq:ode}
\end{equation}
where $I(\alpha)$ is the Fisher information of $\alpha$ under the
tempered target. We follow this approach here, using the assumption
that the sequence of targets is Gaussian. We suppose that $\pi_{0}$
and $\pi_{T}$ are both multivariate normal distributions with common
mean $s_{\text{obs}}$ and covariance $\Sigma_{0}$ and $\varepsilon^{2}\Sigma_{s}$
respectively. Through the argument at the beginning of section \ref{subsec:Choice-of-tempering},
the normality assumption for $\pi_{T}$ is likely to be realistic
when $\varepsilon$ is small. Recall that $\Sigma_{s}$ is a diagonal
matrix chosen to scale the summary statistics. To simplify the following
derivation we assume that $\Sigma_{0}$ is some multiple $\kappa^{2}$
of $\Sigma_{s}$: this is not unreasonable, since in practice a common
choice of $\Sigma_{s}$ is to use the diagonal entries of an estimate
$\hat{\Sigma}_{0}$, obtained through simulation from $f$, of $\Sigma_{0}$
(in this case we would have $\kappa^{2}=1$). Following \citet{chopin_connection_2024}
we obtain that $\pi_{t}(s\mid\theta)$ is a multivariate normal distribution
$\mathcal{N}\left(s\mid s_{\text{obs}},\frac{\varepsilon^{2}\kappa^{2}}{\varepsilon^{2}+\left(\kappa^{2}-\varepsilon^{2}\right)\alpha_{t}}\Sigma_{s}\right)$,
and hence
\begin{eqnarray*}
I(\alpha) & = & -\mathbb{E}\left[\frac{\partial^{2}\log\pi_{t}(s\mid\theta)}{\partial\alpha^{2}}\right]\\
 & = & -\mathbb{E}\left[\frac{\partial^{2}}{\partial\alpha^{2}}\left(-\frac{d_{s}}{2}\log\left(2\pi\right)-\frac{1}{2}\log\det\left(\frac{\varepsilon^{2}\kappa^{2}}{\varepsilon^{2}+\left(\kappa^{2}-\varepsilon^{2}\right)\alpha}\Sigma_{s}\right)-\frac{\varepsilon^{2}+\left(\kappa^{2}-\varepsilon^{2}\right)\alpha}{2\varepsilon^{2}\kappa^{2}}\left(s-s_{\text{obs}}\right)^{T}\Sigma_{s}^{-1}\left(s-s_{\text{obs}}\right)\right)\right]\\
 & = & \frac{1}{2}\mathbb{E}\left[\frac{\partial}{\partial\alpha}\left(\frac{\det\left(\frac{\varepsilon^{2}\kappa^{2}}{\varepsilon^{2}+\left(\kappa^{2}-\varepsilon^{2}\right)\alpha}\Sigma_{s}\right)\text{tr}\left(\left(\frac{\varepsilon^{2}+\left(\kappa^{2}-\varepsilon^{2}\right)\alpha}{\varepsilon^{2}\kappa^{2}}\Sigma_{s}^{-1}\right)\left(\frac{\varepsilon^{2}\kappa^{2}\left(\varepsilon^{2}-\kappa^{2}\right)}{\left(\varepsilon^{2}+\left(\kappa^{2}-\varepsilon^{2}\right)\alpha\right)^{2}}\Sigma_{s}\right)\right)}{\det\left(\frac{\varepsilon^{2}\kappa^{2}}{\varepsilon^{2}+\left(\kappa^{2}-\varepsilon^{2}\right)\alpha}\Sigma_{s}\right)}\right)\right]\\
 &  & \qquad+\frac{1}{2}\mathbb{E}\left[\frac{\partial}{\partial\alpha}\left(\left(s-s_{\text{obs}}\right)^{T}\left(\frac{\varepsilon^{2}+\left(\kappa^{2}-\varepsilon^{2}\right)\alpha}{2\varepsilon^{2}\kappa^{2}}\Sigma_{s}^{-1}+\left(\frac{\varepsilon^{2}+\left(\kappa^{2}-\varepsilon^{2}\right)\alpha}{2\varepsilon^{2}\kappa^{2}}\Sigma_{s}^{-1}\right)^{T}\right)\right)\right]\\
 & = & \frac{1}{2}\mathbb{E}\left[\frac{\partial}{\partial\alpha}\left(\frac{\left(\varepsilon^{2}-\kappa^{2}\right)d_{s}}{\varepsilon^{2}+\left(\kappa^{2}-\varepsilon^{2}\right)\alpha}\right)\right]\\
 &  & \qquad+\mathbb{E}\left[\frac{\partial}{\partial\alpha}\left(\left(s-s_{\text{obs}}\right)^{T}\left(\frac{\varepsilon^{2}+\left(\kappa^{2}-\varepsilon^{2}\right)\alpha}{2\varepsilon^{2}\kappa^{2}}\Sigma_{s}^{-1}\right)\right)\right]\\
 & = & \frac{\left(\varepsilon^{2}-\kappa^{2}\right)^{2}d_{s}}{\left(\varepsilon^{2}+\left(\kappa^{2}-\varepsilon^{2}\right)\alpha\right)^{2}}.
\end{eqnarray*}
The specific instance of equation \ref{eq:ode} is then

\[
\dot{\alpha}=c\left(\frac{\varepsilon^{2}}{\left(\kappa^{2}-\varepsilon^{2}\right)\sqrt{d_{s}}}+\frac{\alpha}{\sqrt{d_{s}}}\right),
\]
whose solution is
\[
\alpha(t)=\exp\left(\frac{ct+k}{\sqrt{d_{s}}}\right)-\frac{\varepsilon^{2}}{\kappa^{2}-\varepsilon^{2}}.
\]
We use this solution to determine a continuous tempering schedule
that has $\alpha(0)=0$ and $\alpha(1)=1$. We obtain
\begin{eqnarray*}
\frac{\varepsilon^{2}}{\kappa^{2}-\varepsilon^{2}} & = & \exp\left(\frac{k}{\sqrt{d_{s}}}\right)\\
\frac{k}{\sqrt{d_{s}}} & = & \log\left(\frac{\varepsilon^{2}}{\kappa^{2}-\varepsilon^{2}}\right)\\
k & = & \sqrt{d_{s}}\log\left(\frac{\varepsilon^{2}}{\kappa^{2}-\varepsilon^{2}}\right)
\end{eqnarray*}
giving
\[
\alpha(t)=\exp\left(\frac{ct+\sqrt{d_{s}}\log\left(\frac{\varepsilon^{2}}{\kappa^{2}-\varepsilon^{2}}\right)}{\sqrt{d_{s}}}\right)-\frac{\varepsilon^{2}}{\kappa^{2}-\varepsilon^{2}},
\]
then
\begin{eqnarray*}
\exp\left(\frac{c+\sqrt{d_{s}}\log\left(\frac{\varepsilon^{2}}{\kappa^{2}-\varepsilon^{2}}\right)}{\sqrt{d_{s}}}\right) & = & 1+\frac{\varepsilon^{2}}{\kappa^{2}-\varepsilon^{2}}\\
c+\sqrt{d_{s}}\log\left(\frac{\varepsilon^{2}}{\kappa^{2}-\varepsilon^{2}}\right) & = & \sqrt{d_{s}}\log\left(1+\frac{\varepsilon^{2}}{\kappa^{2}-\varepsilon^{2}}\right)\\
c & = & \sqrt{d_{s}}\log\left(1+\frac{\varepsilon^{2}}{\kappa^{2}-\varepsilon^{2}}\right)-\sqrt{d_{s}}\log\left(\frac{\varepsilon^{2}}{\kappa^{2}-\varepsilon^{2}}\right)\\
c & = & \sqrt{d_{s}}\left(\log\left(\frac{\kappa^{2}}{\kappa^{2}-\varepsilon^{2}}\right)-\log\left(\frac{\varepsilon^{2}}{\kappa^{2}-\varepsilon^{2}}\right)\right)\\
c & = & \sqrt{d_{s}}\log\left(\frac{\kappa^{2}}{\kappa^{2}-\varepsilon^{2}}\frac{\kappa^{2}-\varepsilon^{2}}{\varepsilon^{2}}\right)\\
c & = & \sqrt{d_{s}}\log\left(\frac{\kappa^{2}}{\varepsilon^{2}}\right)\\
c & = & 2\sqrt{d_{s}}\log\left(\frac{\kappa}{\varepsilon}\right),
\end{eqnarray*}
which gives us
\begin{eqnarray*}
\alpha(t) & = & \exp\left(\frac{2\sqrt{d_{s}}\log\left(\frac{\kappa}{\varepsilon}\right)t+\sqrt{d_{s}}\log\left(\frac{\varepsilon^{2}}{\kappa^{2}-\varepsilon^{2}}\right)}{\sqrt{d_{s}}}\right)-\frac{\varepsilon^{2}}{\kappa^{2}-\varepsilon^{2}}\\
 & = & \exp\left(2\log\left(\frac{\kappa}{\varepsilon}\right)t+\log\left(\frac{\varepsilon^{2}}{\kappa^{2}-\varepsilon^{2}}\right)\right)-\frac{\varepsilon^{2}}{\kappa^{2}-\varepsilon^{2}}.
\end{eqnarray*}

\section{Gaussian marginal likelihood: additional results}

In this section we provide an additional study of the new methods
on the model in section \ref{subsec:Gaussian-marginal-likelihood}.

\begin{figure}
\subfloat[$\log$ RMSE as a function of the number of draws $M$ from $P$.\label{fig:RMSE-as-a-4}]{\includegraphics[scale=0.45]{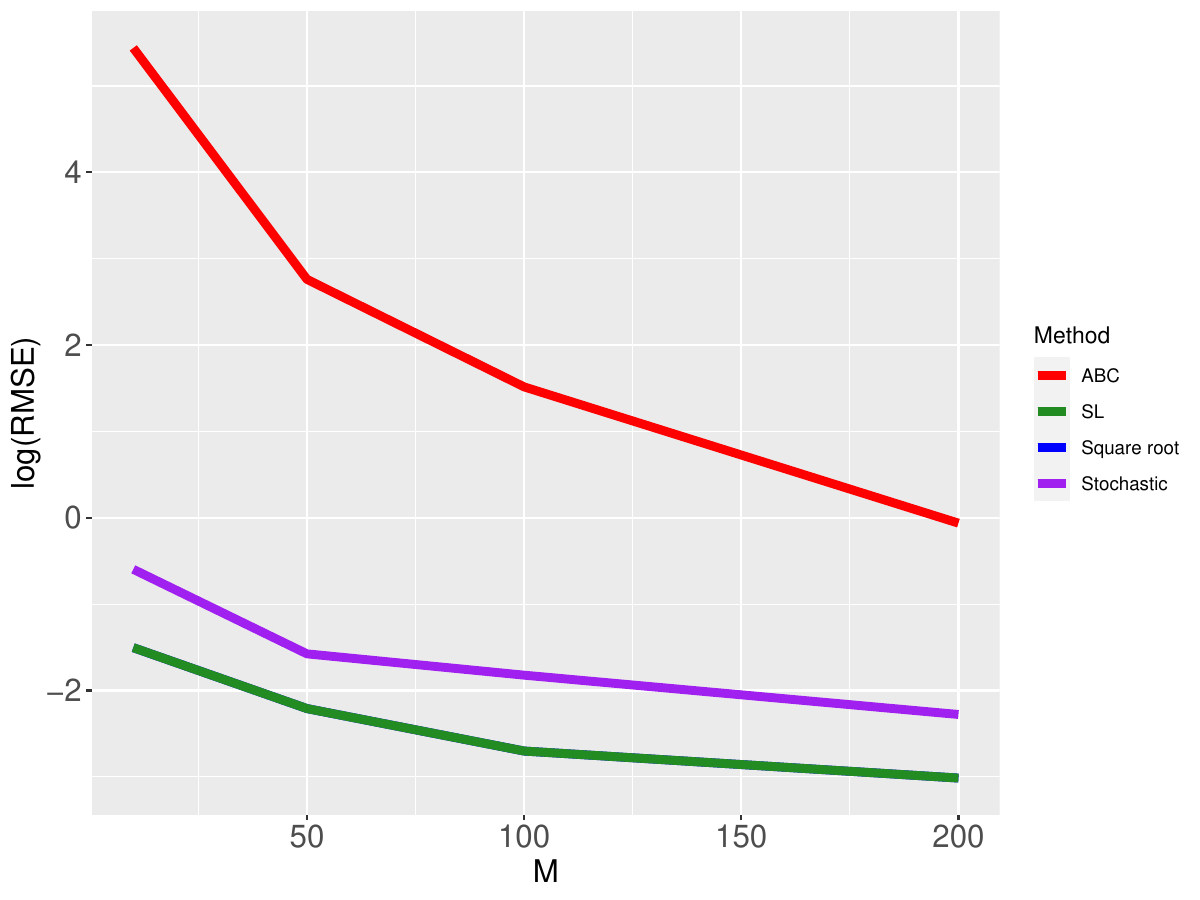}

}\subfloat[$\log$ RMSE of the path sampling estimator as a function of the $\log$
tolerance $\varepsilon$.\label{fig:RMSE-of-the}]{\includegraphics[scale=0.45]{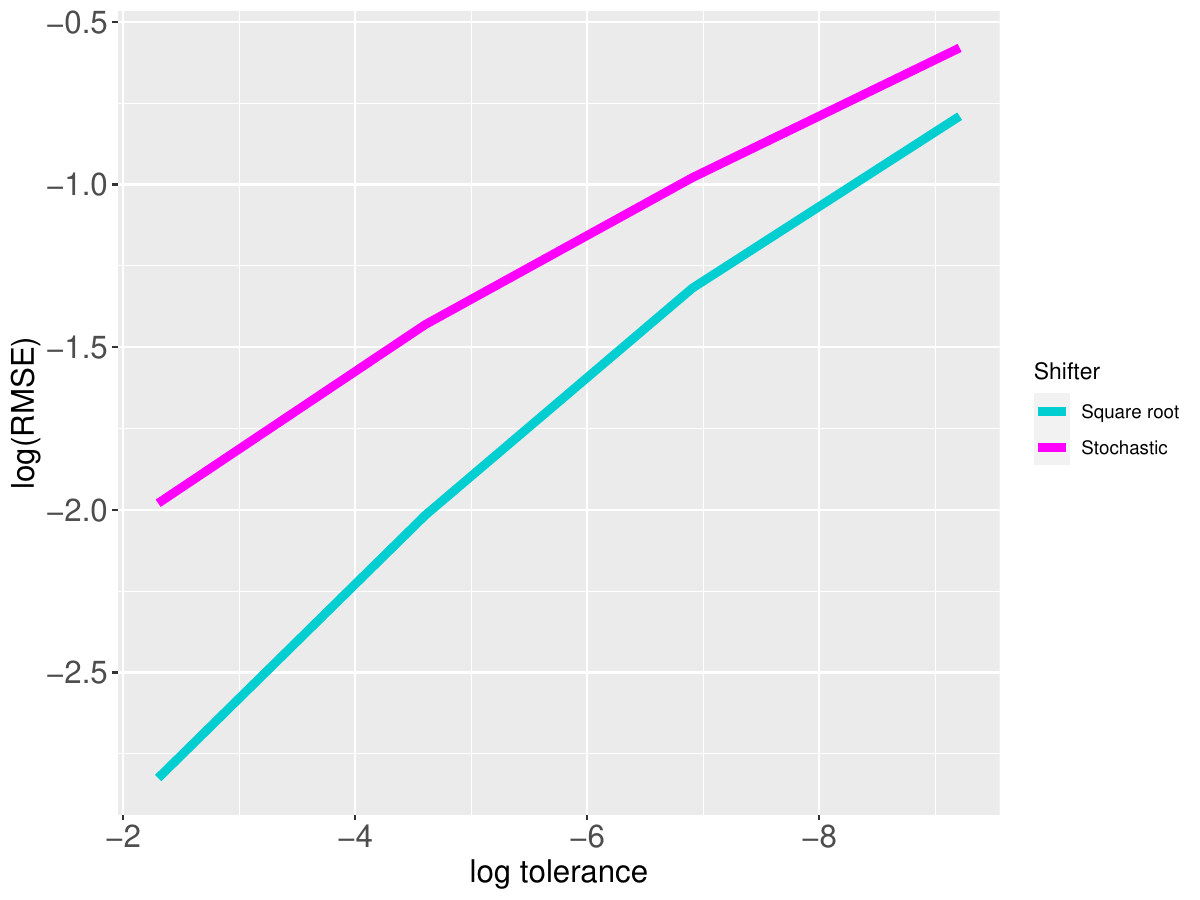}

}

\subfloat[$\log$ RMSE of the path sampling estimator as a function of the number
of targets $T$.\label{fig:RMSE-as-a-2-2}]{\includegraphics[scale=0.45]{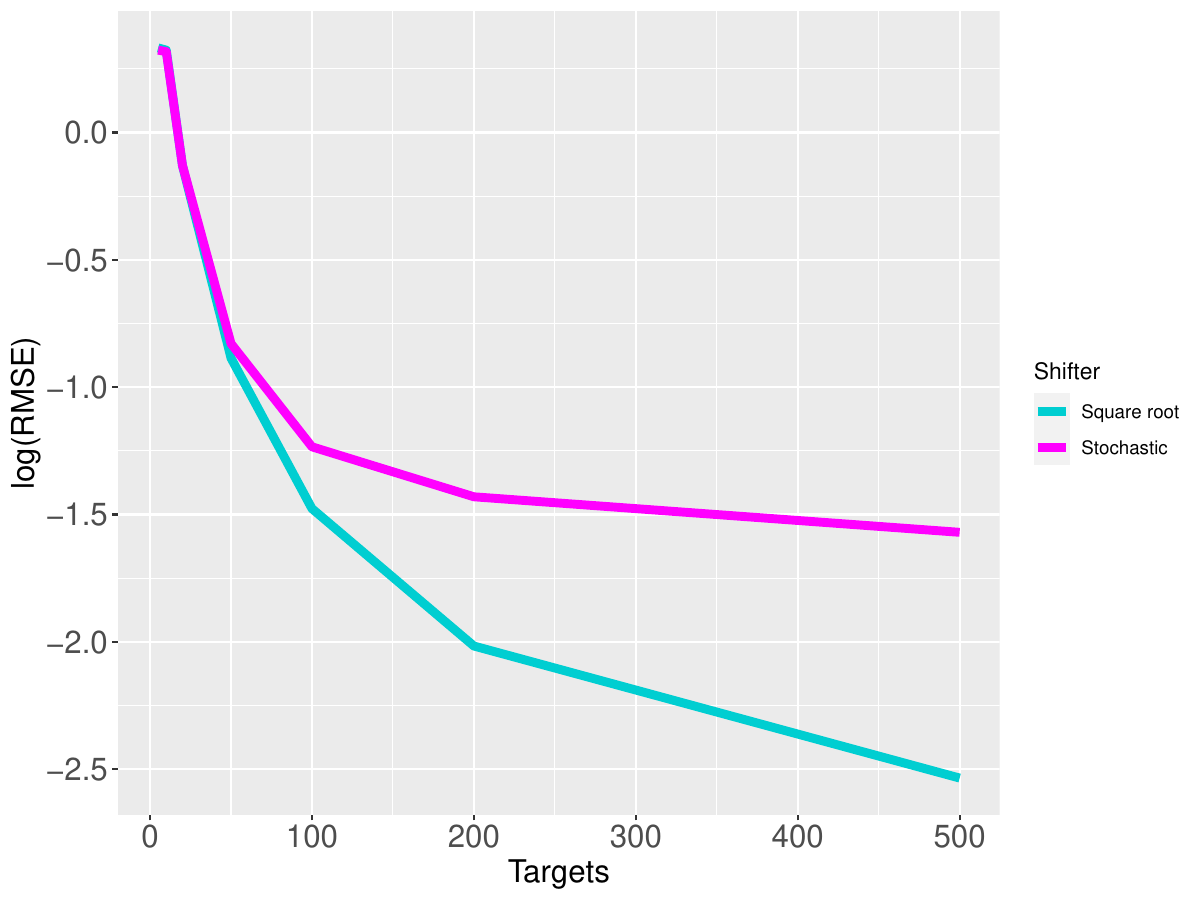}

}\subfloat[MSE times computational time for the path sampling estimator as a
function of $T$.\label{fig:RMSE-of-the-2}]{\includegraphics[scale=0.45]{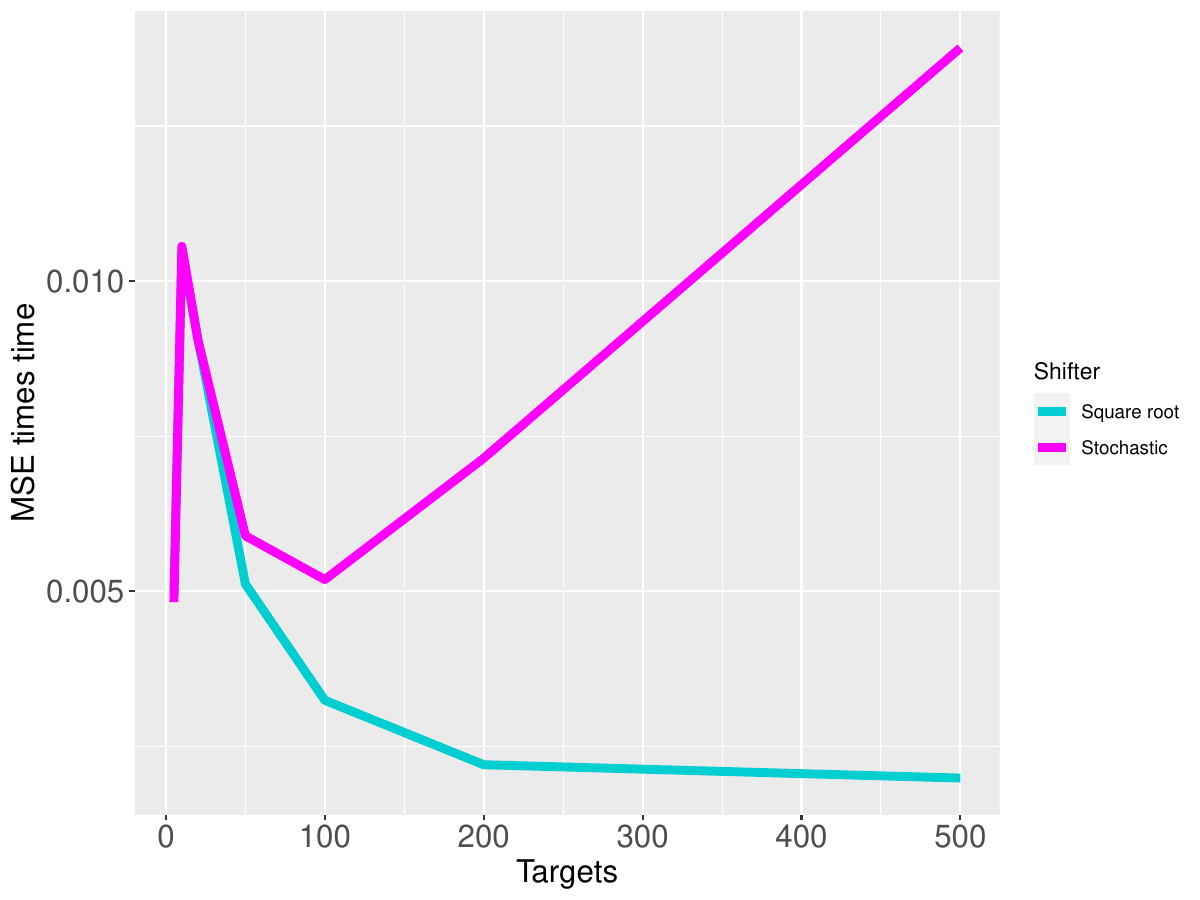}

}

\caption{Properties of IEnKI normalising constant estimators on the Gaussian
example.}
\end{figure}

Figure \ref{fig:RMSE-as-a-4} shows the $\log$ RMSE of the direct
estimator for $\varepsilon=0.01$ as a function of $M$ for ABC, SL and
IEnKI-ABC (where $T=5$) for the two different shifters. The bias
is negligible compared to the variance for this estimator, and we
observe the usual Monte Carlo inverse scaling of the variance with
$M$. ABC is outperformed by the other approaches, and SL has almost
identical performance to IEnKI-ABC with deterministic shifters (the
square root shifter and SL results overlap so only one line can be
seen). The estimator from sIEnKI-ABC has a similar bias to rIEnKI-ABC
and aIEnKI-ABC, but an increased variance due to the noise added in
each update step, as also illustrated in the main text.

For the path sampling estimator we observe the need for an increasing
number of targets to stabilise the RMSE as $\varepsilon$ goes to zero,
reflecting the increased distance between the targets at the start
and end of the sequence. Figure \ref{fig:RMSE-of-the} shows how the
$\log$ estimated RMSE changes with $\log$ $\varepsilon$ (for $T=200$
targets), highlighting the poor scaling of this approach as $\varepsilon$
goes to 0, in contrast with the direct estimator in figure \ref{fig:RMSE-as-a-2}.
Figure \ref{fig:RMSE-as-a-2-2} shows how the $\log$ RMSE of the
path sampling estimator scales with $T$ for $\varepsilon=0.01$, illustrating
the appealing property (compared with the direct estimator in \ref{fig:RMSE-as-a-1})
that an increased $T$ reduces the error, even though this error is
significantly larger than the direct estimator. Figure \ref{fig:RMSE-of-the-2}
shows the estimated $\log$ MSE of the path sampling estimator times
the computational time required to produce the estimate, highlighting
that the additional noise introduced by the stochastic shifter severely
affects the strategy of increasing $T$ to reduce the error.

\foreignlanguage{british}{\bibliographystyle{/Users/Everitt/Dropbox/projects/bib/mychicago}
\bibliography{enki_abc}
}
\end{document}